\documentclass[12pt]{article}
\usepackage{amsmath}
\usepackage{amstext}
\usepackage{amssymb}

\def\al{\alpha}
\def\be{\beta}
\def\ga{\gamma}

\def\ep{\epsilon}
\def\ve{\varepsilon}

\def\et{\eta}
\def\th{\theta}

\def\ka{\kappa}
\def\la{\lambda}

\def\si{\sigma}

\def\ta{\tau}

\def\ch{\chi}
\def\ps{\psi}
\def\om{\omega}
\def\Ga{\Gamma}
\def\De{\Delta}

\def\La{\Lambda}

\def\Om{\Omega}

\def\cl{{\mathcal L}}

\def\fr#1#2{{{#1} \over {#2}}}
\def\half{{\textstyle{1\over 2}}}
\def\frac#1#2{{\textstyle{{#1}\over {#2}}}}
\def\vev#1{\langle {#1}\rangle}

\def\lsim{\mathrel{\rlap{\lower4pt\hbox{\hskip1pt$\sim$}}
   \raise1pt\hbox{$<$}}}
\def\gsim{\mathrel{\rlap{\lower4pt\hbox{\hskip1pt$\sim$}}
   \raise1pt\hbox{$>$}}}
\def\sqr#1#2{{\vcenter{\vbox{\hrule height.#2pt
        \hbox{\vrule width.#2pt height#1pt \kern#1pt
        \vrule width.#2pt}
        \hrule height.#2pt}}}}

\def\prt{\partial}

\def\pt#1{\phantom{#1}}

\newcommand{\beq}{\begin{equation}}
\newcommand{\eeq}{\end{equation}}
\newcommand{\bea}{\begin{eqnarray}}
\newcommand{\eea}{\end{eqnarray}}
\newcommand{\bit}{\begin{itemize}}
\newcommand{\eit}{\end{itemize}}
\newcommand{\rf}[1]{(\ref{#1})}

\begin{document}

\begin{center}
{\Large {\bf Observational Constraints on \\
Local Lorentz Invariance}}\footnote{Chapter to appear in {\it The Springer Handbook of Spacetime}, Springer-Verlag, 2013.}
\\[1cm]
Robert Bluhm\\
Department of Physics\\
Colby College\\
Waterville, ME 04901 USA\\
\end{center}
\vskip 0.2cm

\noindent
\begin{abstract}

The idea that local Lorentz invariance might be violated due to new physics 
that goes beyond the Standard Model of particle physics and Einstein's 
General Relativity has received a great deal of interest in recent years.
At the same time, new experiments have been designed
and conducted that are able to test Lorentz symmetry at unprecedented levels.
Much of this theoretical and experimental progress has been driven by the 
development of the framework for investigating Lorentz violation
known as the Standard Model Extension (SME).
The SME is the lagrangian-based effective field theory that by definition 
contains all Lorentz-violating interaction terms
that can be written as observer scalars involving particle fields in the 
Standard Model and gravitational fields in a generalized theory of gravity.
This includes all terms that could arise from a process of
spontaneous Lorentz violation as well as terms that explicitly break Lorentz symmetry.
In this article, an overview of the SME is presented, 
including its motivations and construction.
A very useful minimal version of the SME in Minkowski spacetime
that maintains gauge invariance and power-counting renormalizability
is constructed as well.
Data tables summarizing tests of local Lorentz invariance for the
different particle sectors in the Standard Model and with gravity
are maintained by Kosteleck\'y's group at Indiana University.
A partial survey of these tests, including some of the high-precision 
sensitivities they attain, is presented here.

\end{abstract}

\pagebreak

\tableofcontents

\newpage

\section{Introduction}

The Standard Model (SM) of particle physics and the theory of General Relativity (GR)
are currently the best theories describing the four fundamental forces of nature:
electromagnetism, the strong and weak nuclear forces, and gravity.
There are no known experimental conflicts with predictions from either of these theories.
Nonetheless, they are fundamentally different in that the SM is a quantum theory,
while GR is a classical geometrical theory.
It remains an open issue as to how to merge or reconcile the SM and GR
into a unified theory that presumably contains a quantum description for gravity.
The relevant scale for a quantum theory of gravity is 
typically taken as the Planck scale,
which is approximately $10^{19}$ GeV.
Promising candidates for a quantum theory of gravity include string theory and loop quantum gravity.
These and other ideas for quantizing gravity can involve new features 
such as, for example, higher dimensions of space and time, brane-world scenarios, noncommutative geometries, 
and spacetime-varying fields or couplings.
It is also possible that in merging GR into a quantum theory of gravity,
the laws of relativity might not hold exactly at all energy scales.

Searching for experimental evidence of a quantum theory of gravity is challenging
because conducting experiments at the Planck scale 
\index{Planck scale}%
is not possible.
However, suppressed effects emerging from a more fundamental theory
might be observable in highly-sensitive low-energy experiments or in interferometry experiments
with extremely long baselines. 
One candidate set of Planck-scale-suppressed signals is relativity violations
associated with small breakings of local Lorentz symmetry.
It has been shown, for example, that mechanisms arising in the context of string theory and
other quantum theories of gravity might lead to violation of Lorentz symmetry.
It is for this reason that considerable interest in the possibility of Lorentz violation
has emerged in recent years, and a number of new high-precision experimental tests of
local Lorentz invariance have been performed.
\index{Lorentz violation}%

A key development in the investigation of Lorentz violation was the formulation of a 
comprehensive theoretical framework known as the Standard-Model Extension (SME)
~\cite{026-kostelecky-95,026-colladay-97,026-colladay-98,026-kostelecky-01a,026-kostelecky-04a}.
\index{Standard-Model Extension (SME)}%
It contains both the SM and generalized theories of gravity (including GR) as well as
all possible observer-independent Lorentz-violating interactions involving particle
and gravitational fields.
The SME has been used extensively to test for Lorentz violation in high-precision experiments.
It also has theoretical features that are important for understanding the different types of
processes that might lead to Lorentz violation.

In this overview, the focus is on using the SME to investigate the possibility of
Lorentz violation both theoretically and experimentally.
An underlying assumption in using the SME is that at low energies
(compared to the Planck scale) Lagrangian-based
field theory gives what is currently
the best description of elementary particles and their interactions.
Therefore, if some new type of physics, such as Lorentz violation,
goes beyond the SM and GR,
then according to this assumption
its leading-order corrections should be describable in
the context of effective field theory.
It is for this reason that the SME is suitable as a framework
for investigating signals of Lorentz violation in experiments.
However, in the search for a consistent quantum theory of gravity,
it is also possible to consider ideas that fall outside the domain
of Lagrangian-based field theory.
These include ideas such as the breakdown of quantum mechanics,
or where spacetime becomes discrete or noncontinuous at the
quantum-gravity scale.
Many of these ideas also lead to Lorentz violation.
To the extent that they can be described at the level of effective field 
theory at low energies,
then they should give rise to effects that are contained in the SME framework.
However, if these alternative theories cannot be described using
Lagrangian-based effective field theory,
then it may not be possible to investigate their effects using the SME.
In those cases, one would need to work within the context of the
given model in order to investigate possible signals of Lorentz violation.

\section{Spacetime Symmetries in Relativity}

In special relativity, the equations of motion for particles
and fields are invariant under Lorentz transformations.
The Lorentz symmetry in this case is a global symmetry,
with the transformations being the same at each point in the spacetime.
The geometry of special relativity is that of a flat spacetime or Minkowski spacetime.  
In contrast, in GR, the effects of gravity are described by the curvature of spacetime,
and the geometry is Riemannian.
Lorentz symmetry still holds in GR, but only locally,
e.g., in instantaneous infinitesimal inertial frames.
In these local frames, 
called local Lorentz frames,
the laws of special relativity are assumed to hold according to the equivalence principle.
The symmetry in this case is local Lorentz invariance (LLI).
\index{local Lorentz invariance (LLI)}%

Curved spacetime in GR is described by the metric tensor, $g_{\mu\nu}$, 
the Riemann curvature tensor, $R^\kappa_{\pt{\kappa}\lambda\mu\nu}$, and its contractions,
including the Ricci tensor, $R_{\mu\nu}$, and the curvature scalar $R$.
These quantities appear in the Einstein field equations along with the energy-momentum
tensor for the matter fields, $T_{\rm M}^{\mu\nu}$, 
which acts as the source of the spacetime curvature.
The Einstein equations are invariant under a set of spacetime transformations 
 (defined mathematically in a later section)
 known as diffeomorphisms, 
 which consist of mappings of the curved spacetime manifold back onto itself.

In particle physics described using special relativity,
the matter fields often have additional symmetries,
such as internal gauge symmetry or
discrete spacetime symmetries.
The latter include parity, P, charge conjugation, C, and time reversal, T.
In the SM of particle physics,
many of these symmetries are broken either explicitly or through a
process of spontaneous symmetry breaking.
For example, spontaneous breaking of gauge symmetry is an essential feature 
of the Higgs mechanism in the electroweak model.
In addition,
all of the discrete symmetries C,P, and T are broken by the weak interactions,
including the combination CP, which is broken in certain meson interactions.
However, a theorem in particle physics, 
known as the CPT theorem, states that the combination of all three
of the discrete spacetime symmetries, CPT,
must hold for all local interactions of point-like particles
in the context of quantum field theory
~\cite{026-schwinger-51,026-bell-55,026-pauli-55,026-luders-57}.
\index{CPT symmetry}%
An essential assumption of the CPT theorem is that Lorentz symmetry must hold.
This is important in investigations of Lorentz violation because it implies
that if Lorentz symmetry is broken, then CPT breaking could occur as well
because the conditions for the theorem to hold would not apply.
This opens up another avenue of investigation of Lorentz violation in that
high-precision tests of CPT symmetry can be used as well to test for Lorentz violation.
Another theorem in the context of quantum field theory 
strengthens this connection. 
It states that in realistic effective field theories,
interactions that break CPT also break Lorentz symmetry
~\cite{026-greenberg-02}. 
Evidently, there is a strong link between CPT violation and Lorentz violation,
and any experiment looking for CPT violation can also be viewed as a Lorentz
test in the context of quantum field theory.

The SM of particle physics can be be combined with GR to describe
all four of the fundamental forces.
This involves using a curved background
with a metric $g_{\mu\nu}$ to describe the physical spacetime
in which the SM particles move and interact.
The resulting theory is a hybrid theory in which the SM fields
are quantum fields with local
$SU(3) \times SU(2) \times U(1)$ gauge symmetry.
However, the metric field is not quantized,
and the pure-gravity sector of the theory remains a classical theory.
A classical Lagrangian can be written down for the full theory
as a sum of a SM sector and a gravity sector,
 \begin{equation}
    \label{eq:026-1}
    {\cal L} = {\cal L}_{\rm SM} + {\cal L}_{\rm GR} \; .
  \end{equation}
 Derivatives of fields included in these expressions must be both
 gauge covariant and gravitationally covariant.
 The latter involves introduction of a spacetime connection $\Gamma^\lambda_{\pt{\lambda}\mu\nu}$. 
 The classical action of the theory is then given by the integral
  \begin{equation}
    \label{eq:026-2}
   S = \int \sqrt{-g}  \, {\cal L} \, d^4 x\; ,
  \end{equation}
  where the factor involving the determinant of the metric, $g$, ensures that the
  spacetime volume element in the integral is covariant under general coordinate transformations.
 The Einstein field equations are obtained by variation
 of the action with respect to the metric.
 The SM fields appear in the Einstein equations by contributing to
 the energy-momentum tensor for the matter fields, $T_{\rm M}^{\mu\nu}$.
Both the action and the field equations are invariant under diffeomorphism transformations.
\index{diffeomorphism transformations}%

In most particle-physics experiments,
the gravitational interactions are irrelevant.
In this case, the contributions from ${\cal L}_{\rm GR}$ are dropped
and the metric is set equal to the Minkowski metric, $\eta_{\mu\nu}$.
The theory can then be treated in the context of quantum field theory,
where special relativity alone suffices.
Without gravity, 
the relevant symmetries are global Lorentz symmetry and the local gauge symmetry of the SM,
 $SU(3) \times SU(2) \times U(1)$.
However, with gravity included, the relevant symmetries of the theory change.
Lorentz symmetry becomes a local symmetry,
and diffeomorphism symmetry becomes important as well.

To observe the LLI of a theory in a curved spacetime,
one approach is to make a coordinate transformation to a local
Lorentz frame at each point in the spacetime manifold.
In this way, the metric in the local coordinate system at each point
reduces to $\eta_{\mu\nu}$,
the connection vanishes,
and locally the laws of special relativity apply.  
The choice of local Lorentz frame is not unique, however,
since a Lorentz transformation at a given point leaves 
$\eta_{\mu\nu}$ unchanged.

An alternative approach keeps the spacetime frame fixed with metric $g_{\mu\nu}$,
but also reveals the LLI at the same time.
In this approach,
four vector fields, $e_\mu^{\pt{\mu}a}$, with labels $a = 0,1,2,3$ are introduced.
They are called vierbein or tetrad fields.
\index{vierbein}%
They relate tensor components in the space-time frame (labeled by
Greek indices) to the corresponding components in a local Lorentz frame 
(labeled by Latin indices).
For example,
the metric obeys
 \begin{equation}
    \label{eq:026-3}
   g_{\mu\nu} = e_\mu^{\pt{\mu}a} e_\nu^{\pt{\nu}b} \eta_{a b} \; .
  \end{equation}
 Since the metric is a symmetric field obeying $g_{\mu\nu} = g_{\nu\mu}$,
 it has at most ten independent degrees of freedom.
 In contrast, the vierbein, $e_\mu^{\pt{\mu}a}$, has a total of sixteen
 independent degrees of freedom.
 The six extra degrees of freedom are associated with the LLI.
 
 There are several advantages to studying possible violations of LLI in a vierbein formalism.
 One is that fermions can more readily be introduced.
 In GR, particles form tensor representations under the group of linear transformations 
associated with general coordinate transformations,
 and there are no representations for spin-half fermions.
 Thus, it is not possible to define Dirac gamma matrices or covariant derivatives of spinor fields
 in a spacetime manifold in GR.
 However, with a vierbein formalism it is possible to extend the usual definitions of
 these quantities in special relativity into curved spacetime.
 Another advantage of a vierbein formalism is that it allows
the local Lorentz symmetry and diffeomorphism symmetry to be treated in a 
manner similar to local gauge symmetry in particle physics.
However, to do this in a general way requires that an additional geometrical 
quantity called torsion be introduced into the theory.
Geometrically, theories with torsion allow a twisting of coordinate axes 
as the axes are transported along a curve.
This twisting cannot be described by the curvature tensor alone.
The resulting geometry when torsion is included is called Riemann-Cartan geometry.
(For reviews describing torsion and Riemann-Cartan geometry, see
\cite{026-hehl-76,026-13-shapiro-76}).
\index{torsion}%
\index{Riemann-Cartan geometry}%
For these reasons,
many investigations of Lorentz violation
use a vierbein formalism
and work in a generalized geometry,
such as Riemann-Cartan geometry.

The use of a vierbein also involves introduction of a spin connection. 
\index{spin connection}%
It enters in covariant derivatives acting on local tensor components and plays the 
role of the gauge field for the Lorentz symmetry. 
In contrast, excitations of the metric field can be viewed as the gauge fields 
for the diffeomorphism symmetry. 
The relationship between the vierbein and spin connection is often a 
reflection of the type of spacetime geometry being considered. 
For example, in a Riemannian geometry (with no torsion), 
the spin connection is nondynamical and does not propagate.
However, in a Riemann-Cartan geometry (with nonzero torsion), 
the spin connection must be treated as independent 
degrees of freedom that in principle can propagate. 
These different types of geometry can have effects on mechanisms 
that occur when Lorentz symmetry is violated. 
This is especially the case when Lorentz symmetry is spontaneously broken.

\subsection{Lorentz Transformations and Diffeomorphisms}

In special relativity, the Lorentz transformations consist of three rotations
and three boosts.
\index{Lorentz transformations}%
They are constant linear transformations that leave the Minkowski metric, $\eta_{\mu\nu}$, invariant.
Mathematically,
they can be implemented by contracting the tensors in a theory with
a transformation matrix $\Lambda_\mu^{\phantom{\mu}\alpha}$.
In Cartesian coordinates, the transformation matrix obeys,
  \begin{equation}
    \label{eq:026-4}
    \eta_{\mu\nu} = \Lambda_\mu^{\phantom{\mu}\alpha} \Lambda_\nu^{\phantom{\nu}\beta} \eta_{\alpha\beta}\; .
  \end{equation}
 
  It is often useful to consider infinitesimal Lorentz transformations,
  which can be written as
  $\Lambda_\mu^{\phantom{\mu}\alpha} \simeq \delta_\mu^{\phantom{\mu}\alpha} + \epsilon_\mu^{\phantom{\mu}\alpha}$,
  where the six parameters, $\epsilon_\mu^{\phantom{\mu}\alpha} = - \epsilon_{\phantom{\alpha}\mu}^{\alpha}$,
  generate infinitesimal rotations and boosts.
  Under an infinitesimal particle Lorentz transformation,
  a tensor $T^{\lambda\mu\nu\cdots}$ transforms as,
   \begin{eqnarray}
    \label{eq:026-5}
   T^{\lambda\mu\nu\cdots} &\rightarrow& T^{\lambda\mu\nu\cdots} 
   + \epsilon^\lambda_{\phantom{\lambda}\rho}T^{\rho\mu\nu\cdots} 
   \nonumber\\ 
   &+& \epsilon^\mu_{\phantom{\mu}\rho} T^{\lambda\rho\nu\cdots}
   + \epsilon^\nu_{\phantom{\nu}\rho} \, T^{\lambda\mu\rho\cdots}
   + \cdots \; .
\end{eqnarray}
  In a theory with LLI,
  the action describing the theory
  and the equations of motion are left unchanged when all of the tensor fields in the
  theory are transformed by infinitesimal Lorentz transformations.

In the presence of gravity,
the vierbein can be used to relate tensor components in a local Lorentz frame
to the corresponding components in the space-time frame.
A vierbein field appears for each tensor index.
For example, for the tensor $T^{\lambda\mu\nu\cdots}$,
 \begin{equation}
    \label{eq:026-6}
   T^{\lambda\mu\nu\cdots} =  e^\lambda_{\pt{\lambda}a}  \,
   e^\mu_{\pt{\mu}b}  \,
   e^\nu_{\pt{\nu}c} \cdots \, T^{abc\cdots}
   + \cdots \; .
  \end{equation}
  A local Lorentz transformation acts on the tensor components
  defined with respect to the local frame,
   e.g., $T^{abc\cdots}$.
  For a local infinitesimal transformation,
  the six Lorentz parameters are written as $\epsilon_{ab}$. 
  These depend on
  the spacetime coordinates at a given point.
Under a local Lorentz transformation,
the vierbein transforms as a vector,
    \begin{equation}
    \label{eq:026-7}
   e_\mu^{\pt{\mu}a}  \rightarrow e_\mu^{\pt{\mu}a} 
   + \epsilon^a_{\phantom{a}d} e_\mu^{\pt{\mu}d} 
   \; .
  \end{equation}
  
  Typically, in a gravitational theory with LLI,
  the six degrees of freedom associated with the local Lorentz symmetry
  are used to gauge away the six anti-symmetric components in the vierbein.
  The remaining ten components are symmetric and can be written
  in terms of field excitations $h_\mu^{\pt{\mu}a} = h_{\pt{a}\mu}^{a}$.
  For small excitations about a flat Minkowski background,
  the form of the vierbein can then be written as
   as
   \begin{equation}
    \label{eq:026-8}
   e_\mu^{\pt{\mu}a} = \delta_\mu^{\pt{\mu}a} 
   + \half h_\mu^{\pt{\mu}a} 
   \; .
  \end{equation}
  Substituting this into Eq.\ \ref{eq:026-3} yields the usual expression for
  the metric in terms of small excitations about a Minkowski background,
  $g_{\mu\nu} = \eta_{\mu\nu} + h_{\mu\nu}$.
  
  Diffeomorphisms are mappings from one differentiable manifold to another.
  In GR, the mappings are from the spacetime manifold back to itself.
  Vectors and tensors transform in prescribed ways under diffeomorphisms,
  and diffeomorphism invariance in GR is the statement that the same physics
  is described by the spacetime manifold, metric, and matter fields both
  before and after a diffeomorphism is performed.
  
  As with local Lorentz symmetry,
  diffeomorphism symmetry can be used to eliminate additional
  degrees of freedom.
  Under infinitesimal diffeomorphism transformations,
  points $x^\mu$ on the space-time manifold are mapped to neighboring points 
  $x^\mu + \xi^\mu$,
  where the four parameters $\xi^\mu$ are spacetime dependent.
  Under infinitesimal diffeomorphisms,
  the metric transforms as
   \begin{equation}
    \label{eq:026-9}
   g_{\mu\nu} \rightarrow g_{\mu\nu} 
   - \partial_\mu \xi_\nu - \partial_\nu \xi_\mu
   \; .
  \end{equation}
  By gauge fixing the four diffeomorphism degrees of freedom, 
  the metric can be reduced 
  from ten down to six independent degrees of freedom.
  The excitations $h_{\mu\nu}$ then have six degrees of freedom
  as well after gauge fixing.
  These represent the six possible excitation modes for gravitational radiation
  that can occur in a generalized theory of gravity.
  For the case of Einstein's GR,
  the kinetic terms in the action are chosen 
  so that four of these degrees of freedom do
  not propagate as physical modes
  and instead are called auxiliary modes.
  As a result, in Einstein's GR only two gravitational modes propagate,
  which are both massless transverse modes.

\subsection{Particle and Observer Transformations}

In investigations of possible Lorentz violation,
it is important to distinguish between {\it observer} and {\it particle} transformations
~\cite{026-colladay-97,026-colladay-98}.
\index{particle transformation}%
\index{observer transformation}%
Observer transformations are essentially changes of coordinate systems,
where the tensors describing particles and fields in the system are left physically unchanged.
Lorentz transformations that transform between different local or global 
inertial frames are examples of observer transformations.
Alternatively, Lorentz transformations can be performed directly on the tensor fields 
in a system, while leaving the observer frame (coordinate system) unchanged.
When performed this way, the transformations are called particle transformations.

Similarly, in GR, general coordinate transformations can be performed,
which correspond to a change of observer frame.
These are observer transformations,
which leave the equations of motion covariant in form.
In contrast, diffeomorphisms are particle transformations performed with respect to a fixed observer,
or in a fixed coordinate frame.
The particles and fields of the system, including for example the metric,
are transformed under diffeomorphisms in a prescribed way
that leaves the equations of motion unchanged.

It is common to hear observer and particle transformations referred to, respectively,
as passive and active transformations.
In theories without spacetime symmetry breaking,
these transformations are essentially inverses of each other
in terms of how they act on tensor quantities. 
However, when a symmetry is broken,
this is no longer the case,
and it is important to make a clearer distinction between 
these two types of transformations.

It is reasonable to assert that a physical interaction should not depend on the 
choice of coordinates of a particular observer.
As a result, any physical theory should be invariant under
the relevant set of observer transformations for that theory.
It is for this reason that observer transformations are not particularly 
meaningful as symmetry transformations.
The physically important symmetry transformations are the particle transformations,
which can be performed in a fixed arbitrary observer frame.

Even in a theory with interactions that break a spacetime symmetry,
the resulting physical description should still not depend on any particular observer
or choice of coordinates.
Thus, in theories with broken Lorentz symmetry,
the Lagrangian and equations of motion should be unchanged
when an observer Lorentz transformation is performed.
However,
with Lorentz-violating interactions,
the particle transformations are no longer symmetries of the theory.
In a given observer frame, the physics can therefore change when a particle or field
is transformed under a particle Lorentz transformation.

For example, consider a scattering experiment in special relativity.
If Lorentz symmetry is violated,
there may be preferred spatial orientations or speeds for the incoming 
and outgoing particles.
As a result, particles scattered in different directions or with different speeds,
with respect to a given observer, may behave differently.
Nonetheless, the theory remains fully observer-independent.
If a different observer measures the same scattering events,
the resulting physical effects will be unaffected.
All that happens in an observer Lorentz transformation is that the same
physical events are expressed with respect to a different Lorentz frame. 

According to this approach,
when LLI is broken, it is only the particle
Lorentz transformations that are broken.
The theory remains Lorentz observer-independent at all times.  
Likewise,
in an extension of GR that incorporates spacetime symmetry breaking,
the relevant transformations are particle local Lorentz transformations
and diffeomorphisms.
If either of these are broken,
the theory should still be covariant under observer local Lorentz
transformations and observer general coordinate transformations.

\subsection{Lorentz Violation}

Lorentz symmetry is fundamental in both the SM and in GR.
It should therefore be tested as accurately as possible as a
way of testing the validity of these theories.
In addition, it has been shown in the context of
quantum-gravity theories that
small violations of Lorentz symmetry might occur.
For example, in string field theory mechanisms can occur that
might lead to spontaneous breaking of Lorentz symmetry
~\cite{026-kostelecky-89a,026-kostelecky-89b,026-kostelecky-89c,
026-kostelecky-91a,026-kostelecky-91b,026-kostelecky-96,026-kostelecky-01b}.
Indeed,
it was this idea that led to the development of the SME,
which in turn has stimulated a variety of new experimental 
tests of LLI.
(For reviews of various experimental
and theoretical approaches to Lorentz and CPT violation see
\cite{026-bluhm-06,026-kostelecky-08a,026-kostelecky-11a,026-liberati-12}).

In string field theory,
\index{string field theory}%
a string state can be expanded as a sum of
tensor-valued particle states,
where the particle masses increase with the order of the tensor.
String interactions provide couplings between the particle states.
Spontaneous Lorentz violation can occur in this context
when a string field theory has a nonperturbative
vacuum that can lead to one or more of the tensor-valued fields, $T$, acquiring nonzero
vacuum expectation values  or vevs, $\langle{T}\rangle \ne 0$.
\index{vacuum expectation value (vev)}%
When this occurs,
the low-energy effective theory can contain terms
of the form 
  \begin{equation}
    \label{eq:026-10}
   \cl \sim {\lambda \over {m_P^k}} \, \langle{T}\rangle \, \Gamma \, \bar \psi (i \partial )^k \chi
   \; ,
  \end{equation}
where $k$ is an integer power, $\lambda$ is a coupling constant,
$\Gamma$ is a generalized Dirac matrix,
$m_P$ is the Planck mass,
and $\psi$ and $\chi$ are fermion fields.
Note that the higher-dimensional ($k>0$) derivative couplings
are expected to be balanced by additional inverse factors of the Planck mass $m_P$. 

In this expression, the tensor vev, $\langle{T}\rangle$, carries spacetime indices,
which are not written out in this notation.
This vev is effectively a set of background functions or constants that are fixed in a given
observer frame.
As tensor-valued backgrounds,
these coefficients can have preferred directions in spacetime or velocity dependence.
In other words,
they induce Lorentz violation.
A more general interaction term can be defined by absorbing all of the couplings
and inverse mass factors into the vev.  
The Lorentz-violating interactions then have the form
  \begin{equation}
    \label{eq:026-11}
   \cl \sim t^{(k)} \, \Ga \, \bar \ps (i \partial )^k \chi
      \; ,
  \end{equation}
where the coefficients $t^{(k)}$
carry spacetime indices
and act as fixed Lorentz-violating background fields.
In addition to interactions with fermions,
additional terms involving gauge-field couplings and gravitational
interactions are possible as well.

The SME is a generalization of these types of interactions
to include all possible contractions of
known SM and gravitational fields 
with fixed background coefficients $t^{(k)}$
~\cite{026-kostelecky-95,026-colladay-97,026-colladay-98,026-kostelecky-01a,026-kostelecky-04a}.
This includes all arbitrary-dimension interaction terms inducing
Lorentz violation in effective field theory.
The coefficients for Lorentz violation, $t^{(k)}$,
are examples of SME coefficients.
\index{SME coefficients}%
They are assumed to be heavily suppressed,
presumably by inverse powers of the Planck mass.
In fact, since no Lorentz violation has been observed in nature,
it must be that these SME coefficients are small.

By developing the SME in this generalized way, 
a framework that is particularly well suited for phenomenology results.
In this approach, 
the Lorentz-violating SME coefficients are treated as
quantities to be bounded in experiments.
They can be thought of as vevs arising
in a process of spontaneous Lorentz violation or simply as being due to
explicit Lorentz violation from some unknown mechanism.

The interactions in \ref{eq:026-10} can also be used to study other
processes related to Lorentz and CPT violation.
For example,
terms of this form have been shown to induce a form of
CPT-violating baryogenesis~\cite{026-bertolami-97}.

Another example of Lorentz violation comes from noncommutative field theory
~\cite{026-connes-98}.
\index{noncommutative field theory}%
These are theories with noncommuting coordinates 
$[x^\mu , x^\nu ] = i \theta^{\mu\nu}$.
It has been shown that this type of geometry can occur naturally
in string theory
and that it leads to Lorentz violation
~\cite{026-mocioiu-00,026-carroll-01,026-guralnik-01,026-carlson-01,026-anisimov-02}.
The fixed parameters $\th^{\mu\nu}$
break Lorentz symmetry and
act effectively as fixed background tensors.
For example,
in an effective field theory with a $U(1)$ gauge field
in a noncommutative geometry, 
interaction terms of the form
 \begin{equation}
    \label{eq:026-12}
  {\cal L} \sim  \, i q \, \theta^{\alpha\beta} \, F_{\alpha \beta} \, \bar \psi
   \, \gamma^\mu \, D_\mu \, \psi 
      \; 
  \end{equation}
can arise.
Here, $F_{\al \be}$ is the field strength,
$q$ is the charge, and $D_\mu$ is a gauge-covariant derivative.
As in Eq.\ \rf{eq:026-12}
the interaction takes the form of a scalar-valued product of
known particle fields, derivative operators, and a set
of fixed background functions.
It is straightforward to write these interactions in
terms of SME couplings.

There are a number of other examples of theories
with Lorentz violation that have been put forward in recent years.
These include models with spacetime-varying fields,
quantum gravity models, multiverses, and braneworld scenarios.
See, for example, 
\cite{026-gambini-99,026-burgess-02,026-alfaro-02,026-amelino-camelia-02,
026-sudarsky-03,026-kostelecky-03,026-frey-03,026-stecker-03,026-bjorken-03,
026-myers-03,026-cline-04,026-mavromotos-04,026-froggatt-05}.
It is also possible to construct models with specific types of Lorentz violation.
These include models that maintain spatial rotational invariance
while breaking only boost transformations,
models with Lorentz-violating dispersion relations constructed
using higher-order derivative interactions,
and vector-tensor models in gravity that 
spontaneously break Lorentz symmetry.
To the extent that these types of models can be described wholly 
or in part using Lagrangian-based effective field they,
they can be investigated using the SME.
However,
some ideas for quantum-gravity theories contain new features
that are not readily described in the context of effective field theory.
Examples include ideas such as spacetime foam,
causal sets, and relative locality.
See, for example,
~\cite{026-christiansen-06,026-christiansen-11,
026-mattingly-08,026-amelino-camelia-11}.
In these types of models, many of the signals of Lorentz violation
that arise are not suitable for investigation using the SME
and instead must be studied in the context of the specific theory.

A number of phenomenological frameworks involving certain kinds of
Lorentz violation have been used by experimentalists in the past.
These include the Robertson-Mansouri-Sexl framework
and the PPN formalism
~\cite{026-robertson-49,026-mansouri-77,026-will-93}.
In some cases,
these and other theories describe parameterized equations of motion 
or dispersion relations that do not originate from a scalar Lagrangian.
However,
to the extent that these models can be described by effective
field theory defined by a scalar Lagrangian,
they are compatible with the SME and direct links between
their parameterizations and the SME coefficients can be obtained.
Since CPT violation in field theory is associated with Lorentz violation,
it follows as well that any observer-independent effective field theory 
describing CPT violation should also be contained within the SME.
Since CPT can be tested to very high precision in experiments
comparing matter and antimatter,
this type of experiment is also ripe as a testing ground
for Lorentz violation.

\section{Standard-Model Extension (SME)}

Currently, there is no consistent quantum
theory of gravity that can be used in detailed examinations of 
the phenomenology of Lorentz violation at accessible energies.
Nonetheless,
progress can still be made using effective field theory.
\index{effective field theory}%
To be realistic,
an effective field theory 
must contain the SM
and a theory of gravity (such as GR),
and it must be compatible with observations.
It must also maintain observer independence.
The Standard-Model Extension (SME) is defined to be
the most general effective field theory of this type incorporating
arbitrary observer-independent Lorentz violation.

The SME lagrangian by definition contains all observer-scalar terms 
that consist of products of SM and gravitational fields with
each other as well as with additional couplings that introduce
violations of Lorentz symmetry.
In principle,
there are an infinity of terms in the SME,
including nonrenormalizable terms of arbitrary dimension.
Most of these terms are expected to be suppressed by
large inverse powers of the Planck scale.
The question of how to extract a useful finite subset of terms
from the full SME to analyze a given experiment becomes relevant,
and there are a number of different ways to proceed.
Perhaps the most natural approach is to follow the direction 
indicated by the experiments testing Lorentz violation.
While these tend to be highly interdisciplinary,
and include experiments in astrophysics, gravity, atomic, nuclear,
and particle physics, 
as well as laboratory experiments with macroscopic media
and space-based tests,
several primary divisions and classifications can be made.

One important split is between experiments that can ignore
the effects of gravity from those that cannot.
For this reason,
a distinction is made between limits of the SME 
that do not include gravity
(where special relativity and global Lorentz invariance are paramount)
from those where gravity is included
(where Lorentz symmetry acts as a local symmetry in a curved spacetime).
It is expected that the nongravitational limits of the SME will in
general be subsets of larger SME limits that include gravity.
For example,
if the curvature is set to zero and the metric is replaced by
the Minkowski metric,
an SME limit with gravity should reduce to a corresponding
SME limit in which gravity is excluded.
Starting from the ground up in constructing
explicit limits of the SME,
it is therefore natural to ignore gravity at first and then
to generalize the resulting theories to incorporate gravity.

In the absence of gravity,
a second primary division between subsets of the SME
can be made based on the types of SM fields and 
interactions (especially their dimensionality) that are included.
Since the SM itself is a renormalizable and gauge-invariant theory,
a first step in constructing a useful SME limit is to
incorporate Lorentz violation while maintaining these features.
This limit restricting the SME to power-counting renormalizable and
gauge-invariant terms is called the minimal SME (mSME).
An advantage of working with the mSME is that each particle 
sector has a finite independent set of mSME
coefficients that can be probed experimentally.
Indeed,
in recent years,
experimentalists have adopted using bounds on mSME coefficients 
as the primary means of reporting sensitivity of their experiments to
Lorentz violation.

Many of the low-energy experiments testing Lorentz 
violation involve only electromagnetic interactions between 
charged particles and photons.
For this reason,
it is useful as well to define a minimal QED sector of the SME.
In a field theory with charged fermions, 
the minimal QED lagrangian consists of the standard Dirac
and Maxwell terms supplemented by Lorentz-violating terms
that maintain U(1) gauge symmetry and power-counting
renormalizability.

If leading order effects are of primary interest,
then SME limits at the level of
relativistic quantum mechanics can be constructed.
This is particularly useful in investigations of low-energy
atomic systems,
where small corrections to atomic energy levels can
result from Lorentz breaking at leading order.
Experiments using particle traps, masers, and 
high-precision spectroscopy can then be analyzed in
a straightforward manner using perturbation theory.

On the other hand,
if first-order effects can be ruled out in an experiment,
it will be necessary to construct limits of the SME that include
nonrenormalizable terms.
In some scenarios for Lorentz violation,
it might happen that Lorentz violation only stems from 
terms of dimension greater than four in the lagrangian.
Alternatively,
if the SME coefficients at leading order are known to
have experimental bounds at levels suppressed by
two powers of the Planck scale,
then it becomes appropriate to look for signals of
Lorentz violation at subleading order as well.
For these reasons,
limits of the SME that contain higher-dimension
nonrenormalizable terms are of interest.

In some experiments,
particular types of particle behaviors play a major
role in attaining sensitivity to Lorentz violation.
Examples include spin-precession effects,
interference, or flavor-changing oscillations.
In these situations it can be advantageous to build
specific types of models out of subsets of the SME.
Such models can then be used as frameworks 
for phenomenology.
It is also useful to consider complementary tests of Lorentz violation
when different experiments only have sensitivity to
combinations of SME experiments.
An example of this involves CPT tests.
These experiments with particles and antiparticles
are typically sensitive at leading
order to combinations of the CPT-odd terms in the SME.
At the same time,
different experiments with the particles alone
might have leading-order sensitivity to different
combinations of both CPT-even and CPT-odd terms.
However,
by analyzing both sets of experiments in terms of
SME coefficients in a complementary manner,
it becomes possible to place more stringent bounds
on individual types of Lorentz violation.

\subsection{Constructing the SME}

The SME contains the SM, a gravity sector, and all possible
observer-independent interactions of these conventional fields with fixed
Lorentz-violating backgrounds,
which are referred to as SME coefficients.
As is typically done in field theory,
the SME can be constructed in terms of a Lagrangian.
The equations of motion are then obtained by variations
of the action with respect to the fundamental fields in the theory.
The SME Lagrangian has three primary sectors,
including one for the SM, one for gravity, and a Lorentz-violating sector,
 \begin{equation}
    \label{eq:026-13}
  {\cal L}_{\rm SME} =  {\cal L}_{\rm SM} + {\cal L}_{\rm GRAV} + {\cal L}_{\rm LV}     \; .
  \end{equation}
The full SME with gravity is defined using a vierbein formalism.
This permits a natural distinction between the spacetime manifold
and local Lorentz frames.

The observer independence of the SME requires that all of the terms in
the Lagrangian be observer scalars under both 
general coordinate transformations and local Lorentz transformations.
This means that every spacetime index 
and every local Lorentz index  
must be fully contracted in the lagrangian.

The SME is not invariant under particle diffeomorphisms and
particle local Lorentz transformations.
The four infinitesimal parameters $\xi^\mu$ 
comprise the diffeomorphism degrees of freedom,
while the six infinitesimal parameters $\ep_{ab} = - \ep_{ba}$ 
carry the six Lorentz degrees of freedom.
In total,
there are ten relevant spacetime symmetries.
Violation of these symmetries occurs when an interaction term 
in the Lagrangian contains SME
coefficients that remain fixed under a particle local Lorentz transformation
or diffeomorphism.

\subsection{Minimal SME}

Since the SM works remarkably well to describe nongravitational particle 
interactions at accessible energies,
it makes sense initially to construct a minimal extension beyond the SM
that contains only those interactions 
for which experiments are likely to have the greatest sensitivity.
These are the interactions that break LLI while maintaining all of
the other desirable features of the SM, 
such as gauge invariance and renormalizability.
The mSME is the restriction of the SME to these power-counting renormalizable and
gauge-invariant terms in the absence of gravity.
\index{minimal SME (mSME)}%

The mSME Lagrangian
can be separated into Lorentz-invariant and Lorentz-
and CPT-violating parts:
 \begin{equation}
    \label{eq:026-14}
{\cal L}_{\rm SME,min} = {\cal L}_{\rm SM} + {\cal L}_{\rm LV,min}    \; .
  \end{equation}
The Lorentz-invariant sector is identified with the
usual Lagrangian for the minimal SM.
The Lorentz-violating Lagrangian 
is the restriction to 
terms of mass dimension 3 and 4 in ${\cal L}_{\rm LV}$
that maintain  
SU(3)$\times$SU(2)$\times$U(1) gauge symmetry.

The first component, ${\cal L}_{\rm SM}$, 
describes the usual interactions for the
strong and electroweak interactions.
The matter fields consist of three generations of quarks and leptons.
These interact through exchange of gauge fields.
A Higgs sector is needed to provide
mass terms for the W and Z bosons in the weak interactions
through the Higgs mechanism,
and Yukawa couplings are needed to give the quarks and
leptons mass terms as well.
The Lagrangian $\cl_{\rm SM}$ can be split into five parts
corresponding to these different sectors:
\begin{eqnarray}
    \label{eq:026-15}
{\cal L}_{\rm SM} &=&
{\cal L}_{\rm lepton} 
+ {\cal L}_{\rm quark} 
+ {\cal L}_{\rm Yukawa} 
\nonumber\\ 
&&
+{\cal L}_{\rm Higgs} 
+ {\cal L}_{\rm gauge}   \; .
\end{eqnarray}
For illustration, the form of the terms
for the lepton sector are given here:
 \begin{equation}
    \label{eq:026-16}
\cl_{\rm lepton} = 
\half i \overline{L}_A \ga^{\mu}\stackrel{\leftrightarrow}{D_\mu} L_A
+ \half i \overline{R}_A \ga^{\mu} \stackrel{\leftrightarrow}{D_\mu} R_A
  \; .
  \end{equation}
In this notation, the left- and right-handed 
lepton multiplets are denoted as
\begin{equation}
    \label{eq:026-17}
L_A = \left( \begin{array}{c} \nu_A \\ l_A
\end{array} \right)_L
\quad , \quad
R_A = (l_A)_R
  \; ,
  \end{equation}
The index $A = 1,2,3$ labels the three flavors,
with $l_A = (e, \mu, \ta)$ denoting the electron, muon, and tau particles,
and $\nu_A = (\nu_e, \nu_\mu, \nu_\ta)$ labeling the three corresponding neutrinos.
The gauge-covariant derivative is denoted $D_\mu$,
and the notation 
$A \stackrel{\leftrightarrow}{\partial_\mu} B 
\equiv A\partial_\mu B - (\partial_\mu A) B$ is adopted.
For the remaining terms in ${\cal L}_{\rm SM} $,
see~\cite{026-colladay-97,026-colladay-98}.

The Lorentz- and CPT-violating part
of the Lagrangian $\cl_{\rm LV,min}$
can also be written as a sum of terms
distinguishing the contributions from the
lepton, quark, Yukawa, Higgs, and gauge sectors.
These partial lagrangians can be further separated 
into CPT-even and CPT-odd parts.
Each of these terms consist of
contractions of the SM fields with SME coefficients.

To illustrate for the lepton sector,
the Lorentz-violating terms are:
\begin{eqnarray}
    \label{eq:026-18}
\cl^{\rm CPT-even}_{\rm lepton} &=&
\half i (c_L)_{\mu\nu AB} \overline{L}_A \ga^{\mu} \stackrel{\leftrightarrow}{D^\nu} L_B
\quad\quad\quad\quad\quad
\nonumber\\ 
&&
+ \half i (c_R)_{\mu\nu AB} \overline{R}_A \ga^{\mu}\stackrel{\leftrightarrow}{D^\nu} R_B  \; ,
\end{eqnarray}
\begin{eqnarray}
    \label{eq:026-19}
\cl^{\rm CPT-odd}_{\rm lepton} &=& 
- (a_L)_{\mu AB} \overline{L}_A \ga^{\mu} L_B
\quad\quad\quad\quad\quad
\nonumber\\ 
&&
- (a_R)_{\mu AB} \overline{R}_A \ga^{\mu} R_B  \; .
\end{eqnarray}
In these expressions,
the SME coefficients $a_\mu$ have dimensions of mass,
while $c_{\mu\nu}$ are dimensionless and traceless.
It is these quantities that act as fixed background fields under
particle Lorentz transformations and induce the breaking of Lorentz symmetry.

\subsection{QED Extension}

The QED  limit of the SME 
\index{QED extension of the SME}%
is useful for specific applications involving
charged particle and photon interactions.
It contains the leading-order
Lorentz- and CPT-violating terms  
that maintain U(1) gauge invariance.
For a single Dirac fermion $\psi$ of mass $m$
the lagrangian is
${\cal L}_{\rm QED, min} 
= {\cal L}_{\rm fermion} + {\cal L}_{\rm photon}$.
The fermion-sector piece can be written as
\begin{equation}
    \label{eq:026-20}
{\cal L}_{\rm fermion}
= {1 \over 2} i \overline{\psi} \Ga^\mu \stackrel{\leftrightarrow}{D_\mu} \psi - \overline{\psi} M \psi  \; ,
  \end{equation}
where the gauge-covariant derivative is
$D_\mu = \prt_\mu + i q A_\mu$
and $\Ga^\nu$ and $M$ are defined by
\begin{eqnarray}
    \label{eq:026-21}
\Ga^\nu &=&
\ga^\nu + c^{\mu\nu} \ga_\mu + d^{\mu\nu} \ga_5 \ga_\mu + e^\nu 
\nonumber\\ 
&&
+ i f^\nu \ga_5 + \half g^{\la\mu\nu} \si_{\la\mu}  \; ,
  \end{eqnarray}
  \begin{eqnarray}
    \label{eq:026-22}
M = m 
+ a_\mu \ga^\mu + b_\mu \ga_5 \ga^\mu 
+ {1 \over 2} H^{\mu\nu} \si_{\mu\nu}  \; .
\end{eqnarray}
  
These equations contain the usual QED terms for a single fermion.
The nonstandard terms violate Lorentz symmetry,
and most have analogues in the minimal SME.
However,
the dimensionless coefficients 
$e^\nu$, $f^\nu$, $g^{\la\mu\nu}$
have no analogue in the minimal SME
because they are incompatible with SU(2)$\times$U(1) symmetry.
They are included in the minimal QED extension
because they are compatible with U(1) invariance
and could emerge from terms in the effective action 
involving the Higgs field.

The lagrangian in the photon sector is  
\begin{eqnarray}
    \label{eq:026-23}
{\cal L}_{\rm photon} &=&
-\frac 14 F_{\mu\nu}F^{\mu\nu}
-\frac 14 (k_F)_{\ka\la\mu\nu} F^{\ka\la} F^{\mu\nu}
\nonumber\\ 
&&
+ {1 \over 2} (k_{AF})^\ka \ep_{\ka\la\mu\nu} A^\la F^{\mu\nu} \; .
  \end{eqnarray}
For simplicity here,
any total-derivative terms are neglected,
as is a possible term of the form $(k_A)_\ka A^\ka$.
Some discussion of the latter can be found in 
Refs.~\cite{026-colladay-97,026-colladay-98}.

In these expressions,
the terms with coefficients $a_\mu$, $b_\mu$,
$e_\mu$, $f_\mu$, $g_{\la\mu\nu}$, and $(k_{AF})_\mu$
are odd under CPT,
while those with
$H_{\mu \nu}$, $c_{\mu \nu}$, $d_{\mu \nu}$, and $(k_F)_{\ka\la\mu\nu}$
preserve CPT.
All ten terms break Lorentz symmetry.
Typically, experiments can have different sensitivities to different
types of Lorentz violation and can involves different particle species.
For this reason,
superscript labels are added to the SME coefficients 
in the fermion sector to denote the particle species.
Lagrangian terms of the same form are expected 
to describe protons and neutrons
in QED systems as well,
but where the SME coefficients represent composites
stemming from quark and gluon interactions.

\subsection{Extensions in Quantum Mechanics}

Many of the sharpest tests of Lorentz symmetry
are  conducted in high-precision particle and atomic experiments.
Typically, static electric and magnetic fields are used in
these experiments to trap or control charged particles,
while the frequencies of particle transitions between
different energy levels are measured with exceptional sensitivity.
The electric and magnetic fields can also be manipulated to 
allow switching between particles and antiparticles,
thereby permitting tests of CPT.
The leading-order shifts in the standard
(Lorentz- and CPT-preserving) energy levels 
are due to the effects of the small quantities
$a_\mu$, $b_\mu$, 
$H_{\mu \nu}$, $c_{\mu \nu}$, $d_{\mu \nu}$,
$e_\mu$, $f_\mu$, and $g_{\la\mu\nu}$.
Lorentz violation stemming from couplings to the
photon coefficients $(k_{AF})_\mu$ and $(k_F)_{\ka\la\mu\nu}$
enters only at subleading order for these types of measurements.

It is often sufficient in calculations describing these systems 
to work at the level of relativistic quantum mechanics using 
a modified Dirac equation.
It is obeyed by a four-component spinor field $\psi$ 
describing a particle with charge $q$ and mass $m$.
Calculation of leading-order energy shifts can be
carried out most readily within a perturbative framework. 
To do so requires extracting a suitable
Dirac hamiltonian from the lagrangian.
However,
the appearance of time-derivative couplings in 
the modified Dirac equation 
means that the standard procedure for obtaining the
Dirac hamiltonian fails to produce 
a hermitian quantum-mechanical operator 
generating time translations on the wave function.
This technical difficulty can be overcome by performing 
a field redefinition at the lagrangian level,
chosen to eliminate the additional time derivatives.
Rewriting the lagrangian in terms of the new field $\chi$
does not affect the physics.
However,
the modified Dirac wave function corresponding to $\chi$
does have conventional time evolution.

The rewritten Dirac equation takes the form 
\begin{equation}
    \label{eq:026-24}
i \partial_0 \ch = \hat H \ch \; ,
  \end{equation}
  with
\begin{equation}
    \label{eq:026-25}
\hat H = \hat H_0 + \hat H_{\rm pert} \; .
  \end{equation}
In this notation, 
$\hat H_0$ is a conventional Dirac hamiltonian 
representing a charged particle 
in the absence of Lorentz- and CPT-violating perturbations.
The perturbative hamiltonian $\hat H_{\rm pert}$ for the particle
is linear in the SME coefficients.
The static electromagnetic fields enter in the perturbative treatment
at leading order only through the dependence of the gauge-covariant 
derivatives on the background potential $A_\mu$.

In many experiments,
energies are probed only at extremely low energy,
where an expansion of the Hamiltonian in a nonrelativistic limit 
is appropriate.
This can be implemented following a 
Foldy-Wouthuysen approach
~\cite{026-kostelecky-99a}.
The resulting nonrelativistic perturbative hamiltonian
can be written in terms of the three-momentum of 
the particle $p_j$ and the usual Pauli matrices $\si^j$
obeying $[\si^j,\si^k]=2i\ve_{jkl}\si^l$.
The leading-order terms are 
\begin{eqnarray}
 \label{eq:026-26}
H_{\rm nonrel} &\simeq& m + {p^2 \over 2 m} 
\quad\quad\quad\quad\quad\quad
\nonumber\\ 
&&
+  ((a_0) - m c_{00}) + 
\nonumber\\ 
&&
+ (-b_j + m d_{j0} \half \epsilon_{jkl} H_{kl})\si^j 
\nonumber\\ 
&&
+ \left[ -a_j + m (c_{oj}+ c_{jo})\right]{p_j \over m} + \cdots .
\end{eqnarray}
In nonrelativistic experiments with ordinary matter 
the primary sensitivity will be to particular combinations
of SME coefficients appearing in these terms.
Subleading contributions can be calculated from
expectation values of the terms involving factors of $p_j$,
where the momentum is treated as a quantum-mechanical
operator.

For experiments designed to test CPT,
which involves measurements of both particles
and antiparticles,
the Dirac hamiltonian for the antiparticle must
also be obtained.
This is accomplished using charge conjugation.
The modified Dirac equation for the antiparticle
differs from that of the particle
by the sign of the charge $q$
and in the sign of any SME coefficients
that are odd under charge conjugation.
See Table 1 for a list of transformation properties
for some of the dominant terms in the QED limit of the mSME.

\begin{table}[t]
\begin{center}
\begin{tabular}{|c|c|c|c|c|c|c|c|}
 \hline
SME Coeff. 	& C 	& P & T & CT & CP & TP & CPT \\ 
 \hline\hline
$a_0$		& $-$ & $+$ & $+$ & $-$ & $-$ & $+$ & $-$ \\
 \hline
$a_j$		& $-$ & $-$ & $-$ & $+$ & $+$ & $+$ & $-$ \\
 \hline
$b_0$		& $+$ & $-$ & $+$ & $+$ & $-$ & $-$ & $-$ \\
 \hline
$b_j$		& $+$ & $+$ & $-$ & $-$ & $+$ & $-$ & $-$ \\
 \hline
$H_{0j}$		& $-$ & $-$ & $+$ & $-$ & $+$ & $-$ & $+$ \\
 \hline
$H_{jk}$		& $-$ & $+$ & $-$ & $+$ & $-$ & $-$ & $+$ \\
 \hline
$c_{00}$		& $+$ & $+$ & $+$ & $+$ & $+$ & $+$ & $+$ \\
 \hline
$c_{0j}$		& $+$ & $-$ & $-$ & $-$ & $-$ & $+$ & $+$ \\
 \hline
 $c_{j0}$		& $+$ & $-$ & $-$ & $-$ & $-$ & $+$ & $+$ \\
 \hline
$c_{jk}$		& $+$ & $+$ & $+$ & $+$ & $+$ & $+$ & $+$ \\
 \hline
$d_{00}$		& $-$ & $-$ & $+$ & $-$ & $+$ & $-$ & $+$ \\
 \hline
$d_{0j}$		& $-$ & $+$ & $-$ & $+$ & $-$ & $-$ & $+$ \\
 \hline
$d_{j0}$		& $-$ & $+$ & $-$ & $+$ & $-$ & $-$ & $+$ \\
 \hline
$d_{jk}$		& $-$ & $-$ & $+$ & $-$ & $+$ & $-$ & $+$ \\
 \hline
\end{tabular}
\end{center}
\begin{center}
{Table 1: Transformation properties of dominant 
SME terms in the matter QED limit under the 
discrete symmetries C,P,T and their combiinations.}
\end{center}
\end{table}

All of the expressions in the quantum-mechanical limits
depend explicitly on the spatial components $j,k,l$ of the 
SME coefficients and on the components 
of various physical quantities,
such as the the particle momenta and the potential $A_\mu$. 
These components are defined 
with respect to a laboratory frame
that must be chosen with a particular orientation.
In laboratory frames fixed with respect to the surface of
the Earth,
the $j=3$ (or $z$ direction) is usually chosen as 
the relevant quantization axis,
typically corresponding to the direction of a static magnetic field.
Alternatively, if a rotation device is used on Earth's surface,
such as a turntable,
its orientation can be chosen as the $j=3$ direction.
In a moving lab,
such as in a satellite orbiting the Earth,
a standard configuration defines the $j=3$ direction
along the satellite velocity with respect to Earth,
with the $j=1$ direction pointing toward Earth
and the $j=2$ direction completing the right-handed system.
In certain situations,
Earth-based experiments may choose to use a satellite-based
configuration as well,
where the velocity of motion is due to the rotation of the Earth
about its axis.
The objective in this case is to take boost effects into account on
the surface of the Earth as is done in satellite experiments.
Ultimately,
no matter which of these alignments is chosen for the
lab frame directions labeled by $j$,
the laboratory axes must be referenced to a
nonrotating basis that can serve as a standard,
since it is only with such a standard basis that comparisons
across different experiments can be made.
Bounds on components of SME coefficients in the lab frame 
must therefore be mapped into bounds on their components 
with respect to the standard reference frame.

For the standard reference frame,
there are a number of different choices that could be made.
Examples include reference frames attached to the centers of mass
of the Earth, the Sun, the Milky Way galaxy, and the cosmic
microwave background radiation (CMBR).
With the exception of the Earth,
each is approximately inertial over thousands of years.
Typically in experiments,
a Sun-centered celestial equatorial frame is chosen
as the standard reference frame.
It is used as the
basis for reporting sensitivities to Lorentz violation.
In certain limits, 
e.g., over short time scales where effects of boosts can be ignored,
the spatial Sun-centered spatial components reduce to
corresponding values in an Earth-based frame.
Similarly, observer transformations from the Sun frame to
a galaxy-based or CMBR-based frame can be made
if bounds are desired with respect to these frames.

\subsection{Gravity Sector}

The gravity sector of the SME uses a vierbein formalism,
which gives the theory a close parallel to gauge theory.
Lorentz breaking occurs due to the presence of SME coefficients,
which remain fixed
under particle Lorentz transformations in a local frame.
In this case,
the SME coefficients carry Latin indices,
e.g., $b_a$ for a vector,
with respect to the local basis set.
The conversion to spacetime coordinates is implemented
by the vierbein,
giving, e.g., $b_\mu = e_\mu^{\phantom{\mu}a} b_a$.
The lagrangian can then be written in terms of fields
and SME coefficients defined on the spacetime manifold.
A natural (though not required) assumption is that
the SME coefficients are smooth functions over the manifold.
It is not necessary to require that they be covariantly constant.
In fact, defining covariantly constant tensors over a manifold
places stringent topological constraints on the geometry.
One simplifying assumption,
which could occur naturally in the context of
spontaneous Lorentz breaking,
is to assume that the SME coefficients are constants in the local frame.
However,
again,
this is not a requirement in the formulation of the SME theory.

To construct the minimal SME including gravity
\cite{026-kostelecky-04a},
the first step is to incorporate gravitational fields into the usual SM.
This is done by rewriting all of the terms in 
the SM Lagrangian with fields and gamma matrices
defined with respect to the local frame (using Latin indices).
The vierbein is then used to convert these terms over to the spacetime manifold.
Factors of the determinant of the vierbein $e$ are included as well
so that integration of the lagrangian density (giving the action)
is covariant.
Derivatives are understood as well to be both spacetime
and gauge covariant.
With these changes,
Eq.\ \rf{eq:026-16}, for example, becomes
\begin{eqnarray}
    \label{eq:026-27}
{\cal L}_{\rm lepton} &=&
{1 \over 2} i e e^\mu_{\phantom{\mu}a}  \overline{L}_A \ga^{a} \stackrel{\leftrightarrow}{D_\mu} L_A
\nonumber\\ 
&&
+ {1 \over 2}i e e^\mu_{\phantom{\mu}a} \overline{R}_A \ga^{a} \stackrel{\leftrightarrow}{D_\mu} R_A \; .
  \end{eqnarray}
The other terms for the quark, Yukawa, Higgs, and gauge
sectors follow a similar pattern.

The Lorentz-violating SME terms constructed from SM fields 
are obtained in a similar way.
The various particle sectors can again be divided between
CPT odd and even contributions.
Each of the terms in the Lagrangian is then written using local indices and vierbeins,
which convert the equations over to the spacetime manifold.
As an example,
Eq.\ \rf{eq:026-18} becomes 
\begin{eqnarray}
    \label{eq:026-28}
{\cal L}^{\rm CPT-even}_{\rm lepton} &=& 
-{1 \over 2} i (c_L)_{\mu\nu AB} e e^\mu_{\phantom{\mu}a} \overline{L}_A \ga^{a} \stackrel{\leftrightarrow}{D^\nu} L_B
\nonumber\\ 
 && - {1 \over 2} i (c_R)_{\mu\nu AB} e e^\mu_{\phantom{\mu}a} \overline{R}_A \ga^{a} \stackrel{\leftrightarrow}{D^\nu} R_B  \; .
\end{eqnarray}
The remaining equations follow the same pattern.

The pure-gravity sector of the minimal SME consists of a
Lorentz-invariant gravity sector and a Lorentz-violating sector.
The Lorentz-invariant lagrangian consists of terms that are products of
the gravitational fields.
In the general case,
this includes terms constructed from curvature, torsion,
and covariant derivatives.
Einstein's gravity (with or without a cosmological term)
would be a special case in this sector.

The Lorentz-violating lagrangian terms in the 
gravity sector of the minimal SME 
are constructed by combining the SME coefficients 
with gravitational field operators
to produce an observer scalar under
local Lorentz transformations
and general coordinate transformations.
These consist of products of the vierbein, the spin connection,
and their derivatives, 
but for simplicity 
they can be written in terms of
the curvature, the torsion $T_{\la\mu\nu}$, and covariant derivatives.
A minimal case (up to dimension four) has the form:
\begin{eqnarray}
    \label{eq:026-29}
{\cl}_{e, \om}^{\rm LV}
&=&
e (k_T)^{\la\mu\nu} T_{\la\mu\nu}
+ e (k_R)^{\ka\la\mu\nu} R_{\ka\la\mu\nu}
\nonumber\\ &&
+ e (k_{TT})^{\al\be\ga\la\mu\nu} T_{\al\be\ga} T_{\la\mu\nu}
\nonumber\\ &&
+ e (k_{DT})^{\ka\la\mu\nu} D_{\ka} T_{\la\mu\nu} .
\end{eqnarray}
The SME coefficients in this expression have 
the symmetries of the associated
Lorentz-violating operators that they multiply.

The Lorentz-violating sector introduces additional gravitational
couplings that can have phenomenological consequences,
including effects on cosmology, black holes, gravitational radiation,
and post-Newtonian physics.
As a starting point for a phenomenological investigation of
the gravitational consequences of Lorentz violation,
it is useful to write down the Riemannian limit of
the minimal SME gravity sector.
It is given as
\cite{026-kostelecky-04a}
\begin{eqnarray}
    \label{eq:026-30}
S_{e, \om, \La}
&=& 
 \fr 1 {2\ka}\int d^4 x 
[ e(1-u)R -2e\La
\quad\quad
\nonumber\\
&&
+ e s^{\mu\nu} R_{\mu\nu}
+ e t^{\ka\la\mu\nu} R_{\ka\la\mu\nu}].
\end{eqnarray}
The SME coefficient $(k_R)^{\ka\la\mu\nu}$ has been expanded into
coefficients $s^{\mu\nu}$, $t^{\ka\la\mu\nu}$, $u$
that distinguish the effects 
involving the Riemann, Ricci, and scalar curvatures.
The coefficients $s^{\mu\nu}$ have the symmetries of the Ricci tensor,
while $t^{\ka\la\mu\nu}$ has those of the Riemann tensor.
Taking tracelessness conditions into account,
there are 19 independent components.

\subsection{Spontaneous Lorentz Violation}

There are a number of theoretical issues concerning Lorentz violation
that can be examined using the SME.
One concerns the nature of the symmetry breaking 
and how that affects the interpretation of the SME coefficients.
These coefficients, e.g., $b_\mu$, for the case of a vector,
couple to SM and gravitational fields as fixed backgrounds.
For the case of a single fermion field, $\psi$, 
in special relativity,
the coupling has the form,
$b_\mu \overline \psi \ga^5 \ga^\mu \psi$.
If this is the only term in the SME Lagrangian containing the
coefficient $b_\mu$,
then the symmetry breaking is said to be explicit.
Essentially the coefficient $b_\mu$ appears in the 
effective field theory without any underlying dynamics.
However, it is also possible for the SME coefficients to arise
through a process of spontaneous symmetry breaking.
In this case,
the SME coefficients are interpreted as vacuum
expectation values (or vevs) of a dynamical tensor field.
For example, for a vector $B_\mu$,
the SME coefficient would arise as a vev, $\vev{B_\mu} = b_\mu$.
The vev acts as a fixed background field that spontaneously
breaks Lorentz symmetry,
but the vector $B_\mu$ remains fully dynamical.

The process of spontaneous symmetry breaking is important
in particle physics.
\index{spontaneous symmetry breaking}%
For example,
in the electroweak theory,
the scalar Higgs field $\phi$ acquires
a nonzero vev, $\vev{\phi} \ne 0$,
that spontaneously breaks the local 
$SU(2) \times U(1)$ gauge symmetry.
For a scalar field,
there is no associated breaking of Lorentz symmetry
because the scalar vev is invariant under Lorentz transformations.
However, the SME coefficients
have tensor indices.
When these occur as nonzero vevs,
then Lorentz symmetry is said to be spontaneously broken.

The standard construction of the SME does not make a
distinction between whether the breaking of Lorentz
symmetry is explicit or spontaneous.
Both types of symmetry breaking can be accommodated,
and both are useful to consider for phenomenological 
investigations of Lorentz violation.
However,
when the gravitational sector of the SME is included,
which brings more geometrical considerations into play,
it becomes important to distinguish these types
of symmetry breaking.

For the case of explicit Lorentz violation,
\index{explicit Lorentz violation}%
it has been shown that inconsistencies arise between 
geometrical constraints  
(e.g., Bianchi identities) and conditions stemming from
the equations of motion.
This was proved by Kosteleck\'y in a no-go theorem
~\cite{026-kostelecky-04a}.
However, the no-go theorem is evaded if the
symmetry breaking is spontaneous.
The crux of the difference has to do with the fact that if
the Lorentz breaking is spontaneous,
then all of the SME coefficients have to be treated as
dynamical fields in field variations.

Because of this, it is often assumed that the SME coefficients
are indeed vevs of dynamical fields that have undergone a
process of spontaneous Lorentz breaking.
Note, however, that if the vevs
are associated with very high energy scales,
then in low-energy tests of Lorentz violation,
they will still act primarily as fixed background fields,
and their dynamics at higher energies will not be relevant.
It is for this reason that the form of the SME or mSME used
by most experimentalists is the same as if the symmetry
breaking were explicit.
For purposes of phenomenology,
the distinction between explicit and spontaneous
Lorentz breaking is not crucial.
For the case of explicit breaking,
it may be that a different type of geometry is relevant,
known as a Riemann-Finsler geometry.
(For a review, see~\cite{026-bao-00}).
The SME with explicit breaking has been shown to be
linked to Riemann-Finsler geometry~\cite{026-kostelecky-11d,026-kostelecky-12b}.

It is certainly the case that spontaneous symmetry breaking is a
very elegant form of symmetry breaking.
This is because when a symmetry is spontaneously broken,
the symmetry still holds dynamically.
However, the vacuum solution for the theory does not obey the symmetry.
What is often done is that a field redefinition is performed
that resets the vacuum values to zero.
In this case, in terms of the new set of fields,
the symmetry becomes hidden at the level of the equations of motion.
It is for this reason that spontaneous symmetry breaking
is also referred to as hidden symmetry.

From a theoretical point of view,
there are well-known consequences when a symmetry
is spontaneously broken.
For example, 
when a global continuous symmetry is spontaneously broken,
it has been shown that massless fields,
called Nambu-Goldstone (NG) fields appear
~\cite{026-nambu-60,026-goldstone-61,026-goldstone-62}.
On the other hand,
if the symmetry is local,
as in the case of the electroweak model,
then a Higgs mechanism can occur
~\cite{026-englert-64,026-higgs-64,026-guralnik-64}.
In this case,
the would-be NG modes get reinterpreted in a way
that results in the gauge fields acquiring a mass.
This is what happens in the electroweak model,
and as a result the W and Z bosons are massive.
However, an unbroken local U(1) gauge symmetry allows the
photons to remain massless.
At the same time,
there are excitations of the Higgs scalar field that
are also massive.
This results in a massive Higgs boson,
which has recently been detected at the Large-Hadron Collider.

An important theoretical issue to consider is whether these
same types of processes can occur when it is Lorentz symmetry
that is spontaneously broken.
\index{spontaneous Lorentz violation}%
For the case where Lorentz symmetry is global,
as in the context of special relativity,
the Goldstone theorem would suggest that massless NG modes should appear.
If so, they would appear as infinite-range particles and would have
implications for phenomenology.
The only known massless particles in the SM and GR
(assuming neutrinos have mass) are the gauge fields,
such as the photon, graviton, and gluons.
Thus, it would seem that there are only two possibilities for the NG modes.
Either the NG modes are known particles,
such as photons or gravitons,
or they are unknown fields that have escaped detection.
However, if the Lorentz symmetry is local,
as in a gravitational theory,
then the question of whether a Higgs mechanism can occur becomes relevant.
In this case,
the possibility of massive gauge fields arises
(massive photons or massive gravity),
and the question of whether there are additional massive Higgs fields
needs to be addressed as well.

These types of questions have been investigated both in special relativity
and in the context of gravity using models that are subsets of the SME.
Interestingly, some of these investigations occurred before the process
of spontaneous symmetry breaking was fully understood.
For example, Dirac worked with a vector model that had a constraint
that the norm of the vector be nonzero
~\cite{026-dirac-51}.
Nambu later showed that such a model spontaneously breaks Lorentz symmetry
~\cite{026-nambu-68}.
Bjorken found a similar model using a composite theory of fermions
that collectively have a nonzero vector vev
~\cite{026-bjorken-63}.
It was conjectured that in these types of models,
the NG modes can be interpreted as photons.
This raises the interesting possibility that photons are massless
because they are the NG modes associated with spontaneous Lorentz breaking,
whereas the conventional idea is that photons are massless because
of local gauge invariance.

In order to impose a constraint that a vector field has a nonzero vev,
the usual process in field theory is to include a potential term that has
a minimum when the vector field equals its vev.
Theories with a vector field and a potential of this type that induces
spontaneous Lorentz violation are known as bumblebee models
~\cite{026-kostelecky-04a,026-kostelecky-89c,026-bluhm-05,026-bluhm-08a,
026-jacobson-01,026-kraus-02,026-moffat-03,026-bertolami-05,026-chkareuli-08,
026-altschul-05,026-bluhm-08b,026-carroll-09,026-seifert-09,026-seifert-10,026-franca-12}.
A defining feature of these theories is that they do not have local
U(1) gauge invariance.
Thus, there is no possibility in these models for photons to arise
because of local U(1) gauge symmetry.
Recent investigations of bumblebee models
have shown that all of the usual processes associated with spontaneous
symmetry breaking can occur when the symmetry is Lorentz symmetry.
First, however, it was found that there is a link between 
local Lorentz symmetry and diffeomorphisms.
In general, if one of these symmetries is spontaneously broken,
then so is the other.
For example, if a vector field has a vev $b_a$ in a local Lorentz frame,
which spontaneously breaks LLI,
then it will also have a vev $b_\mu$ in the spacetime frame,
which spontaneously breaks diffeomorphisms.
(Even for a scalar vev with spacetime dependence this is true,
though in this case it is the derivatives of the scalar that spontaneously
break the symmetries).
What this means is that in the context of a gravitational theory with
spontaneous Lorentz breaking there can be up to ten NG modes,
six associated with Lorentz breaking, and four associated with diffeomorphism breaking.

If these symmetries are treated analogously to local gauge symmetry using
a vierbein formalism,
then it is possible to show that the vierbein itself can accommodate all
ten NG modes when local Lorentz symmetry and diffeomorphisms are
spontaneously broken.
It is also possible to investigate whether a Higgs mechanism can occur
and whether additional massive Higgs modes can appear.
Interestingly, it is found that for a Higgs mechanism to occur the
geometry cannot be Riemannian.
This is because the gauge fields associated with the local Lorentz
symmetry are the spin connection,
and in order to have a dynamical spin connection,
the theory must include torsion.
The geometry must therefore be Riemann-Cartan if a Higgs mechanism 
is to occur.
There can also be additional massive Higgs modes that can affect the
propagation of metric excitations (or gravitational radiation).
It is for this reason that theories of massive gravity often 
result from the process of spontaneous Lorentz violation.
In all of these models, there are stringent conditions that must
hold so that unphysical modes do not appear,
such as negative energy states or tachyons.
These constraints very severely limit the possibilities for making
viable models with massive gravitational fields or massive propagating spin connection.

A subset of the bumblebee models
in which the kinetic term for the vector field has a Maxwell form,
are known as Kosteleck\'y-Samuel (KS) models
~\cite{026-kostelecky-89c}.
For these models,
it has been shown that in the limit where the massive Higgs modes
becomes extremely massive,
the solutions for the KS model match those of Einstein-Maxwell
theory in a fixed gauge.
Thus, the intriguing idea that photons might arise as NG modes
in a theory with spontaneous Lorentz breaking still holds 
even when gravity is included.

It is also possible to consider models with other types of tensor fields
that acquire nonzero vacuum values.
Some possibilities include theories with a symmetric two-tensor
or alternatively an anti-symmetric two-tensor
~\cite{026-kostelecky-05,026-kostelecky-09a,026-altschul-09a}.
Just as with a vector,
when Lorentz symmetry in these models is spontaneously
broken, NG modes and massive modes can appear.
It is useful to study these models to see what the various possibilities are
for the NG and massive modes.
One interesting case is that of a symmetric two-tensor in a Minkowski background.
In this type of model, known as a cardinal model,
the NG modes have properties similar to the graviton
in GR, but in a fixed gauge.
This again raises the intriguing question of whether known
massless particles might occur as a result of Lorentz breaking.
A related consideration is then whether there exist signatures
of the Lorentz breaking that can distinguish KS and cardinal models
from conventional physics.
These types of phenomenological questions can then be suitably addressed
in the context of the SME.

In cosmology, models with spontaneous Lorentz violation have been
used to study modifications of gravity that might give rise to effects
such as accelerated expansion of the universe or to introduce
anisotropic features in the cosmic background radiation.
Examples include
~\cite{026-carroll-04,026-kann0-04,026-lim-05,026-ferreira-07,
026-li-08,026-dulaney-08,026-jimenez-09}.
In general,
these models, which incorporate vector or tensor fields
that spontaneously break Lorentz symmetry,
are studied as possible alternative theories of dark energy.
While these theories have a number of interesting effects
and features,
they do not typically give rise to high-precision 
observational constraints on LLI.
For this reason, these models are not considered here,
and the reader is referred to the literature.

\section{Experimental Tests of Lorentz Violation}

If Lorentz invariance is not an exact symmetry due to mechanisms
occurring in the context of a quantum theory of gravity,
then the relevant energy scale is presumably the Planck scale,
since this is the scale where gravity meets up with quantum physics.
At one time, it was thought unlikely that any physics arising from the
Planck scale would be accessible to experimental detection.
However, with Lorentz violation, the Planck scale is expected to
enter as a suppression factor or inverse power in any corrections
to conventional physics.
Therefore, instead of needing to accelerate particles to
ultra-high energies that are impossible to obtain, 
one can look at extremely high-precision 
experiments often at very low energies for signs of Planck-scale physics. 
In this approach, Lorentz breaking
provides an ideal signal of new physics, 
since nothing in the SM permits violation of Einstein's theory. 
That is, no conventional process could
ever mimic or cover up a genuine signal of Lorentz violation.

The SME serves as a common framework
used by experimentalists and theorists 
to search for signals of Lorentz and CPT violation.
Planck-scale sensitivity has been attained 
to the dominant SME coefficients in a number of experiments
involving different particle sectors.
These include experiments with
mesons, photons, electrons, protons, neutrons,
muons, neutrinos, and the Higgs.
Each particle sector has unique features,
and the experimental methods for testing Lorentz
and CPT violation can differ case by case.

In some experiments,
leading-order sensitivity to Lorentz and CPT violation exists for
more than one particle species at the same time.
This is particularly true in atomic experiments
where bounds involving all three of the
electron, proton, and neutron are often obtained.
Likewise, mixtures of flavors in the meson and neutrino
sectors can occur naturally.
In these cases,
the experimental bounds obtained are for combinations 
of SME coefficients for the different particle sectors.
It is therefore important to look for complementary
sets of bounds obtained from different experiments
that can be combined to select out an optimal
set of bounds for the individual particle species.

In a similar manner,
experiments can have sensitivity to either both CPT-odd and 
CPT-even forms of Lorentz violation,
or alternatively they can probe only the CPT-odd sector.
Most bounds obtained typically involve combinations
of both CPT-odd and CPT-even SME coefficients.
However,
experiments designed to test CPT switch between measurements
on particles and similar measurements on the 
corresponding antiparticles.
The bounds in this case are only on CPT-odd SME coefficients.
For a given particle species,
performing both types of experiments provides a
natural complementary approach.

Before looking at specific experiments,
it is useful to examine some general features that are common to
a number of different experiments.
For example, in low-energy atomic tests,
the sensitivity stems primarily from
the ability of these experiments to detect extremely small anomalous energy shifts.
In many cases, these energy shifts result in small
frequency shifts that can be measured with very high precision.
It is not uncommon for an atomic experiment to be able to measure
a frequency shift with a precision of 1 mHz or less.
Interpreting this as being due to an energy shift expressed in GeV,
it corresponds to a sensitivity of approximately $4 \times 10^{-27}$ GeV.
Such a value is well within the range of energy one
might associate with suppression factors originating
from the Planck scale.
While many of the original atomic experiments 
were designed to measure specific quantities,
such as charge-to-mass ratios of particles and antiparticles
or differences in $g$ factors,
it turns out that it is more effective for these experiments to 
investigate the lowest attainable energy levels for possible anomalous 
shifts associated with Lorentz violation.
Many experiments look specifically for sidereal time
variations of energy levels of a particle or atom
as the Earth moves.
These would result from interactions with the fixed
Lorentz-violating background fields.
Alternatively, experiments designed to test CPT can
 look for instantaneous differences in the energy levels 
of a particle (or atom) and its antiparticle (or antiatom).

Another important general consideration is the choice
of a standard inertial reference frame
~\cite{026-kostelecky-99b}.
Laboratory measurements of Lorentz and CPT symmetry involve
components of SME coefficients defined with respect to a local
laboratory coordinate system.
These components labeled with indices $\{0,j\}$ change
as the lab frame moves or rotates with respect to an inertial frame.
In order to give measured bounds in a consistent manner,
these laboratory bounds must be related to bounds on SME
coefficients defined with respect to a standard inertial frame.
The usual choice for this frame is
a Sun-centered frame that uses 
celestial equatorial coordinates.
Components with respect to the Sun-centered frame
are denoted using upper-case letters
$J,K,L,\dots$ that run over four independent directions
labeled as $\hat T$,$\hat X$,$\hat Y$,$\hat Z$.
The spatial origin of this system is the Sun's center,
and the unit vector $\hat Z$ points along the Earth's
rotation axis, while $\hat X$ and $\hat Y$ lie in the
equatorial plane with $\hat X$ pointing towards the
vernal equinox in the celestial sphere.
The time $T$ is measured by a stationary clock at the origin,
with $T=0$ taken as the vernal equinox in the year 2000.
The Earth's orbital plane lies at an angle $\et \simeq 23^o$
with respect to the $XY$ plane.

Earth-based experiments sensitive to sidereal time 
variations are sensitive to a combination of coefficients,
which are often denoted collectively using tildes.
For example, for electrons,
the combination of spatial components in the lab frame
\begin{equation}
 \label{eq:026-31}
\tilde b_j^e  \equiv b_j^e - m d_{j0}^e - \half \ve_{jkl} H_{kl}^e ,
\end{equation}
arises frequently in a number of experiments.
These combinations are projected onto the nonrotating frame,
where the components  are
$b^e_X$, $b^e_Y$, $b^e_Z$, etc.
Nonrotating frame analogues of the coefficient combinations 
in \ref{eq:026-31} can be defined as
\begin{equation}
 \label{eq:026-32}
\tilde b_J^e  \equiv b_J^e - m d_{J0}^e - \half \ve_{JKL} H_{KL}^e  ,
\end{equation}
where $J,K,L$ label the spatial directions $X,Y,Z$
in the nonrotating frame.
Ignoring boost effects,
the relation between the laboratory and nonrotating 
spatial components is
\begin{eqnarray}
    \label{eq:026-33}
\tilde b_1^e 
&=& 
\tilde b^e_X \cos \ch \cos \Om t
\nonumber\\
&&+ \, \tilde b^e_Y \cos \ch \sin \Om t - \tilde b^e_Z \sin \ch ,
\nonumber\\
\tilde b_2^e 
&=& 
- \tilde b^e_X \sin \Om t + \tilde b^e_Y \cos \Om t ,
\nonumber\\
\tilde b_3^e 
&=& 
\tilde b^e_X \sin \ch \cos \Om t
\nonumber\\
&&+ \, \tilde b^e_Y \sin \ch \sin \Om t + \tilde b^e_Z \cos \ch .
\quad 
\end{eqnarray}
The angle $\ch$ is between the $j=3$ lab axis and the direction
of the Earth's rotation axis along $Z$.
The angular frequency $\Om \simeq {2 \pi}/{({\rm 23 h \, 56 m})}$
is that corresponding to a sidereal day.

\subsection{Data Tables}

A wide range of particle sectors has been investigated for Lorentz and CPT violation.
Many experiments achieve very high sensitivity to Lorentz violation and are
able to place stringent bounds on the relevant SME coefficients.
The results for these bounds are too extensive to list here.
However, a comprehensive summary of Lorentz and CPT tests has been published
by Kosteleck\'y's group at Indiana University~\cite{026-kostelecky-11b}.
It is also updated annually on the physics archive.
\index{Data Tables for Lorentz violation}%

The data tables in Ref.~\cite{026-kostelecky-11b}
provide bounds on Lorentz violation
for ordinary matter (electrons, protons, and neutrons),
photons, mesons, muons, neutrinos, the Higgs, and gravity.
Many tests compare particles and anti-particles.
Low-energy tests in atomic physics include experiments in Penning traps,
comparisons of atomic clocks and masers,
experiments with atomic fountains,
and experiments with antihydrogen at CERN.
Photon tests have been performed using astrophysical and cosmological sources
as well as resonant cavities in the microwave and optical regimes.
Cosmic rays have been investigated for features associated with Lorentz violation.
Experiments with mesons, muons, and neutrinos have used 
large accelerators at high energies.
Experiments are planned or underway on the International Space station,
in space satellites, or using detectors at the south pole.
Experiments with macroscopic torsion pendula 
take advantage of the alignment of large numbers of electron spins to
provide bounds with extremely high sensitivity.
To measure boost effects, some experiments collect data over long periods of
time to enable the Earth's motion to be included.
Other experiments use rotating platforms to gain sensitivity to
a wider range of space-time directions. 

The extremely tight experimental bounds that have been obtained 
on the leading-order SME coefficients indicate that
if Lorentz or CPT violation does occur in nature,
it results in only very small corrections to the SM and GR at ordinary energies.
Since an underlying fundamental theory that would permit
calculation of these corrections is lacking,
at best only order-of-magnitude estimates can be given for the
leading-order SME coefficients.
One possibility is that the leading-order Lorentz-violating terms in the SME
are suppressed by at least one inverse power of the Planck scale.
If a ratio is formed with a low-energy scale on the order of 1 GeV
with the Planck scale,
this results in a suppression factor on the order of $10^{-19}$.
Interestingly,
many of the recent experiments that test Lorentz and CPT symmetry
have sensitivities that are comparable to or exceed expected
order-of-magnitude values based on this suppression factor.
For this reason, it is important as well to search for Lorentz violation
stemming from sub-leading-order terms that are not included in the mSME.
A systematic treatment of these higher-dimensional terms in the SME
has been developed for certain particle sectors,
and bounds on some of these coefficients are included in the data tables as well.

\subsection{Examples}

To highlight some of the Lorentz and CPT tests that have been performed 
a number of different experimental approaches are described here.
In many cases, bounds on a selective subset of SME coefficients are given.
For a full list of experiments with published bounds on SME coefficients,
the reader is referred to the data tables in~\cite{026-kostelecky-11b}.

\medskip
\noindent
$\bullet${\it Penning Traps}~\cite{026-bluhm-97,026-bluhm-98,026-dehmelt-99,026-mittleman-99,026-gabrielse-99}:
\index{Penning traps}%
Experiments in Penning traps use electric and magnetic fields to
isolate and study individual particles and antiparticles.
There are two leading-order signals of Lorentz and CPT violation 
in the electron sector that have been probed in these experiments.
One looks for sidereal time variations in the electron cyclotron and anomaly frequencies.
The idea here is that the Lorentz and CPT-violating interactions depend on
the orientation of the quantization axis in the laboratory frame,
which changes as the Earth turns on its axis.
As a result,
both the cyclotron and anomaly frequencies have small corrections which
cause them to exhibit sidereal time variations.
Such a signal can be measured using just electrons.
Measured bounds are expressed in terms of 
components in the nonrotating Sun-centered frame 
for the combination given in Eq.\ \ref{eq:026-32}.
Their numerical values are on the order of
$|\tilde b_J^e| \lsim  10^{-24} {\rm GeV}$ for $J=X,Y$.
The second type of test in a Penning trap is a traditional CPT test that
compares electrons and positrons directly.
It looks for an instantaneous difference in their
anomaly frequencies.
Leading-order sensitivity in this case involves 
only the CPT-odd coefficient $b^e_3$
(with no tilde),
which is the component of $b^e_\mu$ along the quantization
axis in the laboratory frame.
The bound obtained for $|b^e_3|$ is on the order of $ 10^{-25}$ GeV.

\medskip
\noindent
$\bullet${\it Torsion Pendulum}~\cite{026-bluhm-00a,026-ni-03,026-heckel-06,026-heckel-08}:
\index{torsion pendulum}%
Experiments using a spin-polarized torsion pendulum 
are able to achieve very high sensitivity to
Lorentz violation because the torsion pendulum has a huge
number of aligned electron spins but a negligible magnetic field. 
For example, a pendulum at the University of Washington
is built out of a stack of toroidal magnets,
which has a net electron spin $S \simeq 10^{23}$.
The apparatus is suspended on a rotating turntable and  
the time variations of the twisting pendulum are measured.
An analysis of this system shows that in addition to a signal having the
period of the rotating turntable,
the effects due to Lorentz and CPT violation also cause additional
time variations with a sidereal period caused by the rotation
of the Earth.
Sensitivity to the electron coefficients has been obtained
at the levels of $|\tilde b_J^e| \lsim 10^{-31}$ GeV for $J=X,Y$ and
$|\tilde b_Z^e| \lsim 10^{-30}$ GeV.
By analyzing data over the course of a year, taking the Earth's motion
around the Sun into account,
sensitivity to Lorentz-boost violating coefficients has been attained as well.
This involves a suppression by $v/c \simeq 10^{-4}$,
where $v$ is the velocity of the Earth around the Sun.
The bound on the timelike combination of coefficients is
$\tilde b^e_T \lsim 10^{-27}$ GeV.

\medskip
\noindent
$\bullet${\it Clock-Comparison Tests}~\cite{026-kostelecky-99b,
026-bluhm-99,026-bear-00,
026-phillips-01,026-humphrey-03,026-cane-04,
026-altschul-07,026-gemmel-10,026-altschul-09b,026-smiciklas}:
\index{clock-comparison tests of Lorentz symmetry}%
Many of the sharpest Lorentz bounds for the proton and neutron
stem from atomic clock-comparison experiments.
These involve making high-precision comparisons of 
atomic clock signals as the Earth rotates.
The clock frequencies are typically hyperfine or Zeeman transitions.
Experiments have used hydrogen masers and two-species noble-gas masers 
to achieve the highest sensitivities to Lorentz violation.
For example,
a recent experiment with a K-He$^3$ co-magnetometer
obtained a bound in the neutron sector equal to
$|\tilde b_J^n| \lsim 10^{-33} {\rm GeV}$ for $J=X,Y$.
Experiments with hydrogen masers
attain exceptionally sharp sensitivity to Lorentz and CPT 
violation in the electron and proton sectors.
These experiments use a double-resonance technique that does
not depend on there being a field-independent point for the transition.
The sensitivity for the proton attained in these experiments 
is $|\tilde b_J^p| \lsim 10^{-27}$ GeV.
Due to the simplicity of hydrogen,
this is an extremely clean bound and is one of the more stringent tests
for the proton.
Clock-comparison experiments performed in space 
would have several advantages over traditional
ground-based experiments.
For example,
a clock-comparison experiment conducted aboard 
the International Space Station (ISS)
would be in a laboratory frame that is both rotating and boosted.
It would therefore immediately gain sensitivity to
a wide range of SME coefficients that 
are currently untested~\cite{026-bluhm-02,026-bluhm-03}.
A European mission is planned for
the ISS which will compare atomic clocks and H masers.

\smallskip
\noindent
$\bullet${\it Antihydrogen}~\cite{026-bluhm-99,026-altschul-10,026-fujiwara-11}:
\index{antihydrogen}%
The ALPHA and ATRAP experiments underway at CERN are designed to
produce antihydrogen and to do high-precision spectroscopy on it.
One objective is to
make high-precision spectroscopic measurements of the 1S-2S
transitions in hydrogen and antihydrogen.
These are forbidden (two-photon) transitions that have a relative linewidth
of approximately $10^{-15}$.
The ultimate goal is to measure the line center of this
transition to a part in $10^3$ yielding a frequency comparison
between hydrogen and antihydrogen at a level of $10^{-18}$.
An alternative to 1S-2S transitions is to consider the sensitivity
to Lorentz violation in ground-state Zeeman hyperfine transitions.
It is found that there are leading-order corrections in these levels
in both hydrogen and antihydrogen.
Comparing these measurements for hydrogen and antihydrogen
will provide a direct CPT test.

\smallskip
\noindent
$\bullet${\it Photon Tests}~\cite{026-kostelecky-01c,026-kostelecky-02,026-kostelecky-06,
026-kostelecky-07,026-muller-07,026-klinkhamer-08,026-herrmann-09,
026-eisele-09,026-tobar-09,026-hohensee-10,026-altschul-09c,
026-bocquet-10,026-parker-11}:
\index{photon tests of Lorentz symmetry}%
The relevant leading-order terms for the
photon sector in the SME are the $k_{AF}$
and $k_{F}$ terms in Eq.\ \rf{eq:026-23}.
For the coefficient $k_{AF}$,
which is odd under CPT, 
it is found theoretically that this term leads to negative-energy contributions 
and is a potential source of instability in the theory 
unless it is set to zero~\cite{026-carroll-90}.
In addition, very stringent experimental constraints 
that come from studying the polarization of radiation
from distant radio galaxies also exist
and are consistent with $k_{AF} \approx 0$. 
The terms with coefficients $k_{F}$ are even
under CPT and provide positive-energy contributions.
There are 19 independent components in the
$k_{F}$ coefficients.
Ten of them lead to birefrigence of light.
Bounds on these coefficients of order $10^{-32}$ have been obtained
from spectropolarimetry of light from distant galaxies.
The remaining nine coefficients
have been bounded in a series of laboratory  photon experiments.
These include experiments using optical and microwave cavities,
an Ives-Stilwell experiment, and experiments using rotating platforms.
Sensitivities ranging from $10^{-9}$ up to $10^{-17}$ have been attained
for these coefficients.

\smallskip
\noindent
$\bullet${\it Cosmic Rays}~\cite{026-coleman-99,026-scully-09,026-bi-09}:
\index{cosmic rays}%
Cosmic rays provide the highest-energy particles available experimentally and
can be used to study LLI.
In the presence of Lorentz violation,
the maximal attainable velocity for a cosmic ray in vacuum can
be different from the speed of light by a small amount.  
In principle, it can even exceed the speed of light.  
Effects of this difference include the possibility of
photon decay into electron-positron pairs
or vacuum Cerenkov radiation by ultra-high-energy electrons,
both of which are forbidden in the SM.
Another effect is the prediction in the context of the SM
and special relativity that an upper energy limit
known as the Greisen -- Zatsepin -- Kuzmin, or 
GZK limit~\cite{026-100-greisen-66,026-zatsepin-69},
should hold for cosmic rays emitted from distant sources.  
This theoretical limit is set by interactions with the cosmic
microwave background radiation over long distances.
However, in the presence of Lorentz violation, it is possible
for high-energy cosmic rays from distant sources to exceed the GZK limit.  
This therefore provides an opportunity for testing LLI
and obtaining bounds on the relevant SME coefficients.
Recent experiments at the High Resolution Fly's Eye (HiRes) and 
Pierre Auger Observatory have searched for ultra-high-energy cosmic rays
above the GZK limit,
and their results appear to
confirm the existence of the GZK cutoff.

\smallskip
\noindent
$\bullet${\it Meson Tests}~\cite{026-kostelecky-98,026-kostelecky-00,
026-kostelecky-01d,026-link-03,026-aubert-08,026-didomenico-10,026-kostelecky-10}:
\index{meson tests of Lorentz symmetry}%
Experiments involving neutral meson oscillations
provide very sharp tests of Lorentz and CPT symmetry.
These investigations attain high sensitivity to
the CPT-odd $a_\mu$ coefficients in the SME for
the $K$, $D$, $B_d$, and $B_s$ meson systems.
The time evolution of a meson and its antimeson 
can be described by an effective hamiltonian 
in a description based on the Schr\"odinger equation.
The dominant Lorentz- and CPT-violating contributions to 
the effective hamiltonian can be calculated as expectation 
values of interaction terms in the SME.
The results depend on the velocity of the 
meson with respect to the laboratory frame
and the combinations of SME coefficients $\De a_\mu$,
which vary with sidereal time as the Earth rotates.
Recent analyses have attained bounds on the order of 
$10^{-21}$ GeV for neutral kaons,
$10^{-15}$ GeV in the D system,
$10^{-14}$ GeV for $B_d$ oscillations, 
and $10^{-12}$ GeV for $B_s$ oscillations.

\smallskip
\noindent
$\bullet${\it Muon Tests}~\cite{026-bluhm-00b,026-hughes-01,026-bennett-08}:
\index{muon tests of Lorentz symmetry}%
Lorentz and CPT tests with muons involve second-generation leptons and
are independent of the tests involving electrons.
Several different types of experiments with muons
have been conducted,
including muonium experiments
and $g-2$ experiments with muons.
In muonium,
experiments measuring the frequencies
of ground-state Zeeman hyperfine transitions
in a strong magnetic field have the greatest sensitivity
to Lorentz and CPT violation.
A recent analysis has searched for sidereal time variations
in these transitions.
A bound on SME coefficients, $| \tilde b^\mu_J|$,
has been obtained at a level of $10^{-23}$ GeV.
In relativistic $g-2$ experiments using positive and negative muons
bounds on Lorentz-violation SME coefficients have been obtained at
a level of $10^{-24}$ GeV.

\smallskip
\noindent
$\bullet${\it Collider Tests}~\cite{026-colladay-01,026-hohensee-08,026-abazov-12,026-charneski-12}:	
High energy experiments at colliders provide opportunities for testing Lorentz and CPT violation
in the QED and quark sectors.
Sensitivity for Lorentz violation in cross sections and decay rates has been investigated
in electron-positron scattering.
Effects include variations in observed cross sections with periodicities controlled
by Earth's sidereal rotation frequency.
In a recent experiment using the D0 detector at the Fermilab Tevatron Collider, 
a search for violation of Lorentz invariance in the top quark-antiquark 
production cross section was carried out,
and bounds on SME coefficients for the top quark were obtained.

\smallskip
\noindent
$\bullet${\it Neutrino Tests}~\cite{026-kostelecky-04b,026-kostelecky-04c,026-kostelecky-04d,
026-auerbach-05,026-diaz-09,026-abbasi-10,026-adamson-10,026-diaz-11,026-diaz-12,
026-barger-11,026-katori-11}:
\index{neutrino tests of Lorentz symmetry}%
The experimental observation that neutrinos change flavor when
they propagate through space cannot be explained by the SM.
The conventional explanation for these neutrino oscillations is that
the particles have very small masses.
However, at the same time, the high-precision sensitivity 
of neutrino oscillation experiments, 
stemming from their interferometric nature,
offers possibilities for a range of new tests of LLI.
The neutrino sector of the mSME contains Lorentz-violating
interactions for left-handed neutrinos and right-handed antineutrinos.  
For the left-handed neutrinos,
sensitivity at leading order is to the SME coefficients 
$(a_L)^\mu$ and $(c_L)^{\mu\nu}$.
The resulting signals include ones with the usual $L/E$ dependence,
where $E$ is the energy and $L$ is the oscillation length or baseline 
of the experiment. 
However, with Lorentz violation other dependences, such as
ones with $L$ or $LE$ are possible as well.
These lead to unique signatures of Lorentz violation 
that can occur in neutrino experiments.
These include oscillation, time of flight, and threshold effects.
For example, it has been shown that a Lorentz-violating seesaw mechanism can occur,
which allows for oscillatory behavior even in the absence of mass.
The coefficients for Lorentz violation can also couple to the
four-momentum of the neutrino.
In terrestrial experiments, the direction of the neutrino beam
changes as the Earth rotates,
which leads to sidereal time variations in the oscillation data 
when LLI is broken.
The mSME has been applied to a number of neutrino experiments,
including both short-baseline and long-baseline experiments.  
An extensive list of bounds on SME coefficients in the neutrino
sector are given in the Data Tables~\cite{026-kostelecky-11b}.
For the coefficients $(a_L)^\mu$,
bounds at the level of $10^{-20}$ to $10^{-23}$ GeV have been obtained,
while for the $(c_L)^{\mu\nu}$ coefficients,
the sensitivity ranges from $10^{-17}$ to $10^{-27}$.

\smallskip
\noindent
$\bullet${\it Gravity Tests}~\cite{026-bailey-06,026-battat-07,026-chung-09,
026-muller-08,026-kostelecky-09b,026-kostelecky-11c,026-honesee-11,026-iorio-12}:
\index{gravity tests of Lorentz symmetry}%
Lorentz violation in the gravity sector stems from both
matter-gravity couplings and pure-gravity couplings.
In some cases, the matter-gravity couplings can lead to sensitivity to
forms of Lorentz violation that would otherwise go undetected
in the absence of gravity.
The leading-order SME terms for both these sectors 
in a linearized gravity regime involve expectation 
values denoted as $\bar a_\mu$, $\bar c_{\mu\nu}$
and $\bar s_{\mu\nu}$.
At leading order,
matter-gravity tests are sensitive to $\bar a_\mu$ and $\bar c_{\mu\nu}$,
while pure-gravity tests are sensitive to $\bar s_{\mu\nu}$.
The matter-gravity tests include gravimeter, atom interferometry,
and weak equivalence principle experiments.
Bounds on $\bar a_\mu$ have been obtained at
levels of $10^{-6}$ to $10^{-11}$ GeV
and on $\bar c_{\mu\nu}$ at the 
levels of $10^{-6}$ to $10^{-8}$.
Tests sensitive to the pure-gravity couplings include
experiments with atom interferometers, torsion pendula,
and lunar and satellite laser ranging experiments.
Bounds on $\bar s_{\mu\nu}$ coefficients 
at levels of $10^{-6}$ to $10^{-9}$ have been obtained.  
In addition to these gravity tests,
highly sensitive tests attempting to detect spacetime torsion can be achieved by
searching for its couplings to fermions~\cite{026-kostelecky-08b}.
Bounds on torsion components down to levels of $10^{-31}$ GeV have
been obtained in this way.

\section{Summary and Conclusions}

Interest in the idea of Lorentz violation has 
steadily increased over the past two decades.
This is due to theoretical advances showing that 
Lorentz breaking can provide unique signals of 
Planck-scale physics and quantum-gravity effects
as well as to experimental advances that have led to
new high-precision tests of LLI.
The development and use of the SME
as the theoretical framework describing Lorentz violation in the
context of field theory has led to a comprehensive
and multi-disciplinary approach to testing LLI that
spans most of the particle sectors in the SM.

The underlying premise of the SME is that field theory and
the SM are correct descriptions of particle interactions at low energies.
Therefore, any indications of Lorentz violation should
show up as small corrections in the context of effective field theory.
The SME is constructed as the most general effective field
theory that incorporates Lorentz violation.
It contains all known particle fields and gravitational interactions
as well as all observer-independent terms that break LLI.
As an incremental first step,
the minimal SME and its QED limit,
which maintain gauge invariance and power-counting
renormalizability,
were constructed in the 1990s.
These have been used extensively to search for leading-order
signals of Lorentz and CPT violation.
More recently, a systematic approach to constructing the nonminimal
sectors of the SME have been worked out for certain particle species,
and experimental bounds of these terms are being obtained 
as well~\cite{026-kostelecky-09c,026-kostelecky-12a}.  

As a comprehensive theoretical framework,
the SME allows for investigations of theoretical issues
related to the idea of Lorentz violation.
Specifically, for the case of spontaneous Lorentz breaking,
investigations of the fate of the Nambu-Goldstone modes 
and the possibility of Higgs masses and a Higgs mechanism
have been carried out.
It has been shown that spontaneous Lorentz violation 
is accompanied by spontaneous diffeomorphism breaking,
and up to 10 Nambu-Goldstone modes can appear in principle.
These modes can comprise 10 of the 16 degrees of freedom of the vierbein,
which in a Lorentz-invariant theory are 
gauge degrees of freedom.
The fate of the Nambu-Goldstone modes
is found to depend on the type of spacetime geometry in the
underlying theory.
At leading order in Minkowski and Riemann spacetimes,
it is found that the Nambu-Goldstone modes can propagate like photons 
in a fixed axial gauge.
However, in Riemann-Cartan spacetimes,
the possibility exists 
that the spin connection can absorb the Nambu-Goldstone modes 
in a gravitational version of the Higgs mechanism.
In addition, the potential inducing spontaneous Lorentz violation
can provide mass terms for the metric excitations.
These features create new possibilities for constructing models
with spontaneous Lorentz violation in the context of massive gravity.

\begin{table}[t]
\begin{center}
\noindent
\renewcommand{\arraystretch}{1.2}
\begin{tabular}{|c|c|c|c|}
\hline\hline
Expt & Sector & Params ($J=X,Y)$ & Bound (GeV)
\\
\hline\hline
Penning Trap & electron & $\tilde b_J^e$ & $10^{-24}$ \\[2mm]
\cline{1-4}
K-He dual maser & neutron & $\tilde b_J^n$ & $ 10^{-33}$ \\[2mm]
\cline{1-4}
H maser & proton & $\tilde b_J^p$ & $10^{-27}$ \\[2mm]
\cline{1-4}
Muonium & muon & $\tilde b_J^\mu$ & $10^{-23}$ \\[2mm]
\cline{1-4}
Spin Pendulum & electron & $\tilde b_J^e$ & $10^{-31}$ \\[2mm]
\hline
\hline
\end{tabular}
\end{center}
\renewcommand{\arraystretch}{1.0}
\begin{center}
{Table 2: Summary of leading-order bounds for the coefficient $\tilde b_J$.}
\vspace{0.2cm}
\end{center}
\end{table}

The main application of the SME has been in phenomenological
investigations of Lorentz and CPT symmetry.
High precision tests have been performed in most of the
primary particle sectors in the SM.  
These include experiments in QED and atomic systems,
astrophysical tests, and laboratory tests at nuclear and
particle facilities.  
The generality of the SME allows comparisons across different types
of experiments involving the same particle species.
These tests have greatly improved the sensitivity to which
Lorentz and CPT symmetry is known to hold,
though many particle sectors, particularly those beyond leading order,
remain to be probed.  
As a comparison of some of the bounds obtained to date at leading order,
a summary of some bounds on $\tilde b_J$ coefficients
in the minimal SME is given in Table 2.
These bounds are within the range of sensitivity associated
with suppression factors arising from the Planck scale.
A more complete set of tables for the full SME is published in the
Indiana University data tables~\cite{026-kostelecky-11b}.


\newpage

\begin{thebibliography}{0}

\bibitem{026-kostelecky-95}
V.A.\ Kosteleck\'y and R.\ Potting,
Phys.\ Rev.\ D {\bf 51}, 3923 (1995).

\bibitem{026-colladay-97}
D.\ Colladay and V.A.\ Kosteleck\'y,
Phys.\ Rev.\ D {\bf 55}, 6760 (1997).

\bibitem{026-colladay-98}
D.\ Colladay and V.A.\ Kosteleck\'y,
Phys.\ Rev.\ D {\bf 58}, 116002 (1998).

\bibitem{026-kostelecky-01a}
V.A.\ Kosteleck\'y and R.\ Lehnert,
Phys.\ Rev.\ D {\bf 63}, 065008 (2001).

\bibitem{026-kostelecky-04a}
V.A.\ Kosteleck\'y,
Phys.\ Rev.\ D {\bf 69}, 105009 (2004).

\bibitem{026-schwinger-51}
J. Schwinger,
Phys. Rev. {\bf 82} (1951) 914.

\bibitem{026-bell-55}
J.S. Bell,
Proc. Roy. Soc. (London) {\bf A 231} (1955) 479.

\bibitem{026-pauli-55}
W. Pauli,
in W. Pauli, ed.,
{\it Neils Bohr and the Development of Physics},
McGraw-Hill, New York, 1955, p. 30.

\bibitem{026-luders-57}
G. L\"uders,
Ann. Phys. (N.Y.) {\bf 2} (1957) 1.

\bibitem{026-greenberg-02}
O.W.\ Greenberg,
Phys.\ Rev.\ Lett.\ {\bf 89}, 231602 (2002).

\bibitem{026-hehl-76}
F.W.\ Hehl {\it et al.},
Rev.\ Mod.\ Phys.\ {\bf 48}, 393 (1976).

\bibitem{026-13-shapiro-76}
I.L.\ Shapiro,
Phys.\ Rep.\ {\bf 357}, 113 (2002).

\bibitem{026-kostelecky-89a}
V.A.\ Kosteleck\'y and S.\ Samuel,
Phys.\ Rev.\ D {\bf 39}, 683 (1989).

\bibitem{026-kostelecky-89b}
V.A.\ Kosteleck\'y and S.\ Samuel,
Phys.\ Rev.\ Lett.\ {\bf 63}, 224 (1989).

\bibitem{026-kostelecky-89c}
V.A.\ Kosteleck\'y and S.\ Samuel,
Phys.\ Rev.\ D {\bf 40}, 1886 (1989).

\bibitem{026-kostelecky-91a}
V.A.\ Kosteleck\'y and R.\ Potting,
Nucl.\ Phys.\ B {\bf 359}, 545 (1991).

\bibitem{026-kostelecky-91b}
V.A.\ Kosteleck\'y and S.\ Samuel,
Phys.\ Rev.\ Lett.\ {\bf 66}, 1811 (1991).

\bibitem{026-kostelecky-96}
V.A.\ Kosteleck\'y and R.\ Potting,
Phys.\ Lett.\ B {\bf 381}, 89 (1996).

\bibitem{026-kostelecky-01b}
V.A.\ Kosteleck\'y and R.\ Potting,
Phys.\ Rev.\ D {\bf 63}, 046007 (2001).

\bibitem{026-bluhm-06}
R.\ Bluhm,
Lect.\ Notes Phys.\ {\bf 702}, 191 (2006).

\bibitem{026-kostelecky-08a}
V.A.\ Kosteleck\'y, ed.,
{\it CPT and Lorentz Symmetry IV},
World Scientific, Singapore, 2008.

\bibitem{026-kostelecky-11a}
V.A.\ Kosteleck\'y, ed.,
{\it CPT and Lorentz Symmetry V}
(World Scientific, Singapore, 2011).

\bibitem{026-liberati-12}
S. Liberati and D. Mattingly,
arXiv:1208.1071.

\bibitem{026-bertolami-97}
O.\ Bertolami {\it et al.},
Phys.\ Lett.\ B {\bf 395}, 178 (1997).

\bibitem{026-connes-98}
A.\ Connes, M.\ Douglas, and A.\ Schwartz,
J.\ High Energy Phys.\ {\bf 02}, 003 (1998).

\bibitem{026-mocioiu-00}
I.\ Mocioiu, M.\ Pospelov, and R.\ Roiban,
Phys.\ Lett.\ B {\bf 489}, 390 (2000).

\bibitem{026-carroll-01}
S.M.\ Carroll {\it et al.},
Phys.\ Rev.\ Lett.\ {\bf 87}, 141601 (2001).

\bibitem{026-guralnik-01}
Z.\ Guralnik, R.\ Jackiw, S.Y.\ Pi, and A.P.\ Polychronakos,
Phys.\ Lett.\ B {\bf 517}, 450 (2001).

\bibitem{026-carlson-01}
C.E.\ Carlson, C.D.\ Carone, and R.F.\ Lebed,
Phys.\ Lett.\ B {\bf 518}, 201 (2001).

\bibitem{026-anisimov-02}
A.\ Anisimov, T.\ Banks, M.\ Dine, and M.\ Graesser,
Phys.\ Rev.\ D {\bf 65}, 085032 (2002).

\bibitem{026-gambini-99}
R.\ Gambini and J.\ Pullin,
Phys.\ Rev.\ D {\bf 59}, 124021 (1999).

\bibitem{026-burgess-02}
C.P.\ Burgess, J.\ Cline, E.\ Filotas,
J.\ Matias, and G.D.\ Moore,
JHEP {\bf 0203}, 043 (2002).

\bibitem{026-alfaro-02}
J.\ Alfaro, H.A.\ Morales-T\'ecotl, and L.F.\ Urrutia,
Phys.\ Rev.\ D {\bf 66}, 124006 (2002).

\bibitem{026-amelino-camelia-02}
G.\ Amelino-Camelia,
Mod.\ Phys.\ Lett.\ A {\bf 17}, 899 (2002).

\bibitem{026-sudarsky-03}
D.\ Sudarsky, L.\ Urrutia, and H.\ Vucetich,
Phys.\ Rev.\ D {\bf 68}, 024010 (2003).

\bibitem{026-kostelecky-03}
V.A.\ Kosteleck\'y, R.\ Lehnert, and M.\ Perry,
Phys.\ Rev.\ D {\bf 68}, 123511 (2003).

\bibitem{026-frey-03}
A.R.\ Frey,
JHEP {\bf 0304}, 012 (2003).

\bibitem{026-stecker-03}
F.W.\ Stecker,
Astropart.\ Phys.\ {\bf 20}, 85 (2003).

\bibitem{026-bjorken-03}
J.D.\ Bjorken,
Phys.\ Rev.\ D {\bf 67}, 043508 (2003).

\bibitem{026-myers-03}
R.\ Myers and M.\ Pospelov,
Phys.\ Rev.\ Lett.\ {\bf 90}, 211601 (2003).

\bibitem{026-cline-04}
J.\ Cline and L.\ Valc\'arcel,
JHEP {\bf 0403}, 032 (2004).

\bibitem{026-mavromotos-04}
N.E.\ Mavromatos,
Nucl.\ Instrum.\ Meth.\ B {\bf 214}, 1 (2004).

\bibitem{026-froggatt-05}
C.D.\ Froggatt and H.B.\ Nielsen,
Ann.\ Phys.\ (Leipzig) 14, 115 (2005).

\bibitem{026-christiansen-06}
W.A.\ Christiansen, Y.J.\ Ng, and H.\ van Dam,
Phys.\ Rev.\ Lett.\ {\bf 96}, 051301 (2006).

\bibitem{026-christiansen-11}
W.A.\ Christiansen, Y.J.\ Ng, D.J.E.\ Floyd, and E.S.\ Perlman,
Phys.\ Rev.\ D {\bf 83}, 084003 (2011).

\bibitem{026-mattingly-08}
D.\ Mattingly,
Phys.\ Rev.\ D {\bf 77}, 125021 (2008).

\bibitem{026-amelino-camelia-11}
G.\ Amelino-Camelia, L.\ Freidel, J.\ Kowalski-Glikman, and L.\ Smolin,
Phys.\ Rev.\ D {\bf 84}, 084010 (2011).

\bibitem{026-robertson-49}
H.P.\ Robertson, Rev.\ Mod.\ Phys.\ {\bf 21}, 378 (1949).

\bibitem{026-mansouri-77}
R.\ Mansouri and R.U.\ Sexl, Gen.\ Rel. Grav.\ {\bf 8}, 497 (1977).

\bibitem{026-will-93}
C.N.\ Will,
{\it Theory and experimentation in Gravitational Physics}
(Cambridge University Press, Cambridge, England, 1993).

\bibitem{026-kostelecky-99a}
V.A.\ Kosteleck\'y and C.D.\ Lane,
J.\ Math.\ Phys.\ {\bf 40}, 6245 (1999).

\bibitem{026-bao-00}
D.\ Bao, S.-S.\ Chern, and Z. Shen,
{\it An Introduction to Riemann-Finsler Geometry}
(Springer, New York, 2000).

\bibitem{026-kostelecky-11d}
V.A.\ Kosteleck\'y, 
Phys.\ Lett.\ B {\bf 701}, 470 (2011).

\bibitem{026-kostelecky-12b}
V.A.\ Kosteleck\'y, N.\ Russell, and R.\ Tso,
Phys.\ Lett.\ B {\bf 716}, 470 (2012).
	
\bibitem{026-nambu-60}
Y.\ Nambu,
Phys.\ Rev.\ Lett.\ {\bf 4}, 380 (1960).

\bibitem{026-goldstone-61}
J.\ Goldstone,
Nuov.\ Cim.\ {\bf 19}, 154 (1961).

\bibitem{026-goldstone-62}
J.\ Goldstone, A.\ Salam, and S.\ Weinberg,
Phys.\ Rev.\ {\bf 127}, 965 (1962).

\bibitem{026-englert-64}
F.\ Englert and R.\ Brout,
Phys.\ Rev.\ Lett.\ {\bf 13}, 321 (1964).

\bibitem{026-higgs-64}
P.W.\ Higgs,
Phys.\ Rev.\ Lett.\ {\bf 13}, 508 (1964).

\bibitem{026-guralnik-64}
G.S.\ Guralnik, C.R.\ Hagen, and T.W.B.\ Kibble,
Phys.\ Rev.\ Lett.\ {\bf 13}, 585 (1964).

\bibitem{026-dirac-51}
P.A.M.\ Dirac,
Proc.\ R.\ Soc.\ Lon.\ {\bf A209}, 291, (1951).

\bibitem{026-nambu-68}
Y.\ Nambu,
Prog.\ Theor.\ Phys.\ Suppl.\ Extra 190 (1968).

\bibitem{026-bjorken-63}
J.D.\ Bjorken,
Ann.\ Phys.\ {\bf 24}, 174 (1963).

\bibitem{026-bluhm-05}
R.\ Bluhm and V.A.\ Kosteleck\'y,
Phys.\ Rev.\ D {\bf 71}, 065008 (2005).

\bibitem{026-bluhm-08a}
R.\ Bluhm, S.-H.\ Fung, and V.A.\ Kosteleck\'y,
Phys.\ Rev.\ D {\bf 77}, 065020 (2008).

\bibitem{026-jacobson-01}
T.\ Jacobson and D.\ Mattingly,
Phys.\ Rev.\ D {\bf 64}, 024028 (2001).

\bibitem{026-kraus-02}
P.\ Kraus and E.T.\ Tomboulis,
Phys.\ Rev.\ D {\bf 66}, 045015 (2002).

\bibitem{026-moffat-03}
J.W.\ Moffat,
Intl.\ J.\ Mod.\ Phys.\ D {\bf 12}, 1279 (2003).

\bibitem{026-bertolami-05}
O.\ Bertolami and J.\ Paramos, 
Phys.\ Rev.\ D {\bf 72}, 044001 (2005).

\bibitem{026-altschul-05}
B.\ Altschul, and V.A.\ Kosteleck\'y,
Phys.\ Lett.\ B {\bf 628}, 106 (2005).

\bibitem{026-chkareuli-08}
J.L.\ Chkareuli {\it et al.},
Nucl.\ Phys.\ B {\bf 796}, 211 (2008).

\bibitem{026-bluhm-08b}
R.\ Bluhm, N.\ Gagne, R.\ Potting, and A. Vrublevskis,
Phys.\ Rev.\ D {\bf 77}, 125007 (2008).

\bibitem{026-carroll-09}
S.M.\ Carroll, T.R.\ Dulaney, M.I.\ Gresham, and H.\ Tam,
Phys.\ Rev.\ D {\bf 79}, 065011 (2009).

\bibitem{026-seifert-09}
M.D.\ Seifert,
Phys.\ Rev.\ D {\bf 79}, 124012 (2009).

\bibitem{026-seifert-10}
M.D.\ Seifert,
Phys.\ Rev.\ D {\bf 81}, 065010 (2010).

\bibitem{026-franca-12}
O.J.\ Franca, R.\ Montemajor, L.F. Urrutia,
Phys.\ Rev.\ D {\bf 85}, 085008 (2012). 

\bibitem{026-kostelecky-05}
V.A.\ Kosteleck\'y and R.\ Potting,
Gen.\ Rel.\ Grav.\ {\bf 37}, 1675 (2005)

\bibitem{026-kostelecky-09a}
V.A.\ Kosteleck\'y and R.\ Potting,
Phys.\ Rev.\ D {\bf 79}, 065018 (2009).

\bibitem{026-altschul-09a}
B.\ Altschul, Q.G.\ Bailey, and V.A.\ Kosteleck\'y,
Phys.\ Rev.\ D {\bf 81}, 065028 (2010).

\bibitem{026-carroll-04}
S.M.\ Carroll and E.A.\ Lim,
Phys.\ Rev.\ D {\bf 70}, 123525 (2004).

\bibitem{026-kann0-04}
S.\ Kanno and J.\ Soda,
Phys.\ Rev.\ D {\bf 74}, 063505 (2006).

\bibitem{026-lim-05}
E.A.\ Lim,
Phys.\ Rev.\ D {\bf 71}, 063504 (2005).

\bibitem{026-ferreira-07}
P.G. Ferreira, B.M.\ Gripaios, and R.\ Saffari,
Phys.\ Rev.\ D {\bf 75}, 044014 (2007).

\bibitem{026-li-08}
B.\ Li, D.F.\ Mota, and J.D.\ Barrow,
Phys.\ Rev.\ D {\bf 77}, 024032 (2008).

\bibitem{026-dulaney-08}
T.R.\ Dulaney, M.I.\ Gresham, and M.B.\ Wise,
Phys.\ Rev.\ D {\bf 77}, 083510 (2008).

\bibitem{026-jimenez-09}
J.\ Beltran Jimenez and A.L.\ Maroto,
Phys.\ Rev.\ D {\bf 80}, 063513 (2009).

\bibitem{026-kostelecky-99b}
V.A.\ Kosteleck\'y and C.D.\ Lane,
Phys.\ Rev.\ D {\bf 60}, 116010 (1999);

\bibitem{026-kostelecky-11b}
V.A.\ Kosteleck\'y, and N.\ Russell,
Rev.\ Mod.\ Phys.\ {\bf 83}, 11 (2011);
arXiv:0801.0287.

\bibitem{026-bluhm-97}
R.\ Bluhm, V.A.\ Kosteleck\'y, and N.\ Russell,
Phys.\ Rev.\ Lett.\ {\bf 79}, 1432 (1997).

\bibitem{026-bluhm-98}
R.\ Bluhm, V.A.\ Kosteleck\'y, and N.\ Russell,
Phys.\ Rev.\ D {\bf 57}, 3932 (1998).

\bibitem{026-dehmelt-99}
H.\ Dehmelt {\it et al.}.
Phys.\ Rev.\ Lett.\ {\bf 83}, 4694 (1999).

\bibitem{026-mittleman-99}
R.\ Mittleman {\it et al.},
Phys.\ Rev.\ Lett.\ {\bf 83}, 2116 (1999).

\bibitem{026-gabrielse-99}
G.\ Gabrielse {\it et al.}.
Phys.\ Rev.\ Lett.\ {\bf 82}, 3198 (1999).

\bibitem{026-bluhm-00a}
R.\ Bluhm and V.A.\ Kosteleck\'y,
Phys.\ Rev.\ Lett.\ {\bf 84}, 1381 (2000).

\bibitem{026-ni-03}
L.-S.\ Hou, W.-T.\ Ni, and Y.-C.M.\ Li,
Phys.\ Rev.\ Lett.\ {\bf 90}, 201101 (2003).

\bibitem{026-heckel-06}
B.\ Heckel {\it et al.},
Phys.\ Rev.\ Lett.\ {\bf 97}, 021603 (2006).

\bibitem{026-heckel-08}
B.\ Heckel {\it et al.}l,
Phys.\ Rev.\ D {\bf 78}, 092006 (2008).

\bibitem{026-bluhm-99}
R.\ Bluhm, V.A.\ Kosteleck\'y, and N.\ Russell,
Phys.\ Rev.\ Lett.\ {\bf 82}, 2254 (1999).

\bibitem{026-bear-00}
D.\ Bear {\it et al.},
Phys.\ Rev.\ Lett.\ {\bf 85}, 5038 (2000).

\bibitem{026-phillips-01}
D.F.\ Phillips {\it et al.},
Phys.\ Rev.\ D {\bf 63}, 111101 (2001).

\bibitem{026-humphrey-03}
M.A.\ Humphrey {\it et al.},
Phys.\ Rev.\ A {\bf 68}, 063807 (2003).

\bibitem{026-cane-04}
F.\ Can\`e {\it et al.},
Phys.\ Rev.\ Lett.\ {\bf 93}, 230801 (2004).

\bibitem{026-altschul-07}
B.\ Altschul,
Phys.\ Rev.\ D {\bf 75}, 041301 (2007).

\bibitem{026-brown-10}
J.M.\ Brown {\it et al.},
Phys.\ Rev.\ Lett.\ {\bf 105}, 151604 (2010).

\bibitem{026-gemmel-10}
C.\ Gemmel {\it et al.},
Phys.\ Rev.\ D {\bf 82}, 111901 (2010).

\bibitem{026-altschul-09b}
B.\ Altschul,
Phys.\ Rev.\ D {\bf 79}, 061702 (2009).

\bibitem{026-smiciklas}
M.\ Smiciklas {\it et al.},
Phys.\ Rev.\ Lett.\ {\bf 107}, 171604 (2011).

\bibitem{026-bluhm-02}
R.\ Bluhm {\it et al.},
Phys.\ Rev.\ Lett.\ {\bf 88}, 090801 (2002).

\bibitem{026-bluhm-03}
R.\ Bluhm {\it et al.},
Phys.\ Rev.\ D {\bf 68}, 125008 (2003).

\bibitem{026-altschul-10}
B.\ Altschul,
Phys.\ Rev.\ D {\bf 81}, 041701 (2010).

\bibitem{026-fujiwara-11}
M.C.\ Fujiwara {\it et al.}, ALPHA collaboration, in
V.A.\ Kosteleck\'y, ed.,
{\it CPT and Lorentz Symmetry V}
(World Scientific, Singapore, 2011).

\bibitem{026-kostelecky-01c}
V.A.\ Kosteleck\'y and M.\ Mewes,
Phys.\ Rev.\ Lett.\ {\bf 87}, 251304 (2001).

\bibitem{026-kostelecky-02}
V.A.\ Kosteleck\'y and M.\ Mewes,
Phys.\ Rev.\ D {\bf 66}, 056005 (2002).

\bibitem{026-kostelecky-06}
V.A.\ Kosteleck\'y and M.\ Mewes,
Phys.\ Rev.\ Lett.\ {\bf 97}, 140401 (2006).

\bibitem{026-kostelecky-07}
V.A.\ Kosteleck\'y and M.\ Mewes,
Phys.\ Rev.\ Lett.\ {\bf 99}, 011601 (2007).

\bibitem{026-muller-07}
H.\ M\"uller {\it et al.},
Phys.\ Rev.\ Lett.\ {\bf 99}, 050401 (2007).

\bibitem{026-klinkhamer-08}
F.R.\ Klinkhamer and M.\ Risse,
Phys.\ Rev.\ D {\bf 77}, 117901 (2008).

\bibitem{026-herrmann-09}
S.\ Hermann {\it et al.},
Phys.\ Rev.\ D {\bf 80}, 105011 (2009).

\bibitem{026-eisele-09}
Ch.\ Eisele, A.\ Yu.\ Nevsky, and S.\ Schiller,
Phys.\ Rev.\ Lett.\ {\bf 103}, 090401 (2009).

\bibitem{026-tobar-09}
M.E.\ Tobar {\it et al.}, 
Phys.\ Rev.\ D {\bf 80}, 125024 (2009).

\bibitem{026-hohensee-10}
M.A.\ Hohensee {\it et al.},
Phys.\ Rev.\ D {\bf 82}, 076001 (2010).

\bibitem{026-altschul-09c}
B.\ Altschul,
Phys.\ Rev.\ D {\bf 80}, 091901 (2009).

\bibitem{026-bocquet-10}
J.-P.\ Bocquet {\it et al.},
Phys.\ Rev.\ Lett.\ {\bf 104}, 241601 (2010).

\bibitem{026-parker-11}
S.\ Parker {\it et al.},
Phys.\ Rev.\ Lett.\ {\bf 106}, 180401 (2011).

\bibitem{026-carroll-90}
S.M.\ Carroll, G.B.\ Field, and R.\ Jackiw,
Phys.\ Rev.\ D {\bf 41}, 1231 (1990).

\bibitem{026-coleman-99}
S.\ Coleman and S.L.\ Glashow,
Phys.\ Rev.\ D {\bf 59}, 116008 (1999).

\bibitem{026-scully-09}
S.T.\ Scully and F.W.\ Stecker,
Astroparticle Phys.\ {\bf 31}, 220 (2009).

\bibitem{026-bi-09} 
X.-J.\ Bi {\it et al.},
Phys.\ Rev.\ D {\bf 79}, 083015 (2009).

\bibitem{026-100-greisen-66}
K. Greisen, Phys. Rev. Lett. {\bf 16}, 748 (1966). 

\bibitem{026-zatsepin-69}
G.T. Zatsepin, V.A. KuzÕmin, in Cosmic rays, 
Moscow, No. 11, p. 45 - 47, vol. 11 (1969),
vol. 11, pp. 45Ð47.

\bibitem{026-kostelecky-98}
V.A.\ Kosteleck\'y,
Phys.\ Rev.\ Lett.\ {\bf 80}, 1818 (1998);

\bibitem{026-kostelecky-00}
V.A.\ Kosteleck\'y,
Phys.\ Rev.\ D {\bf 61}, 016002 (2000);

\bibitem{026-kostelecky-01d}
V.A.\ Kosteleck\'y,
Phys.\ Rev.\ D {\bf 64}, 076001 (2001).

\bibitem{026-link-03}
J.M.\ Link {\it et al.}, FOCUS Collaboration,
Phys.\ Lett.\ B {\bf 556}, 7 (2003).

\bibitem{026-aubert-08}
B.\ Aubert
{\it et al.},
BaBar Collaboration,
Phys.\ Rev.\ Lett.\ {\bf 100}, 131802 (2008).

\bibitem{026-didomenico-10}
A.\ Di Domenico, KLOE collaboration,
Found.\ Phys.\ {\bf 40}, 852 (2010).

\bibitem{026-kostelecky-10}
V.A.\ Kosteleck\'y and R.J.\ Van Kooten,
Phys.\ Rev.\ D {\bf 82}, 101702 (2010).

\bibitem{026-bluhm-00b}
R. Bluhm, V.A.\ Kosteleck\'y and C.D. Lane,
Phys. Rev. Lett. {\bf 84}, 1098 (2000).

\bibitem{026-hughes-01}
V.W.\ Hughes {\it et al.},
Phys. Rev. Lett. {\bf 87}, 111804 (2001).

\bibitem{026-bennett-08}
G.W.\ Bennett {\it et al.},
Phys.\ Rev.\ Lett.\ {\bf 100} 091602 (2008).

\bibitem{026-colladay-01}
D.\ Colladay and V.A.\ Kosteleck\'y,
Phys.\ Lett.\ B {\bf 511}, 209 (2001).

\bibitem{026-hohensee-08}
M.A.\ Hohensee {\it et al.}, 
Phys.\ Rev.\ D {\bf 80}, 036010 (2009).
	
\bibitem{026-abazov-12}
V.M.\ Abazov {\it et al.} (D0 Collaboration), 
Phys.\ Rev.\ Lett.\ {\bf 108}, 261603 (2012).

\bibitem{026-charneski-12}
B.\ Charneski {\it et al.}, 
Phys.\ Rev.\ D {\bf 86}, 045003 (2012).

\bibitem{026-kostelecky-04b}
V.A.\ Kosteleck\'y and M.\ Mewes,
Phys.\ Rev.\ D {\bf 69}, 016005 (2004).

\bibitem{026-kostelecky-04c}
V.A.\ Kosteleck\'y and M.\ Mewes,
Phys.\ Rev.\ D {\bf 70}, 031902(R) (2004).

\bibitem{026-kostelecky-04d}
V.A.\ Kosteleck\'y and M.\ Mewes,
Phys.\ Rev.\ D {\bf 70}, 076002 (2004).

\bibitem{026-auerbach-05}
L.B.\ Auerbach {\it et al.}, LSND Collaboration
Phys.\ Rev.\ D {\bf 72} 076004 (2005).

\bibitem{026-diaz-09}
J.S.\ Diaz, V.A.\ Kosteleck\'y and M.\ Mewes,
Phys.\ Rev.\ D {\bf 80}, 076007 (2009).

\bibitem{026-abbasi-10}
R.\ Abbasi {\it et al.}, IceCube Collaboration
Phys.\ Rev.\ D {\bf 82} 112003 (2010).

\bibitem{026-adamson-10}
P.\ Adamson {\it et al.}, MINOS Collaboration
Phys.\ Rev.\ Lett.\ {\bf 105} 151601 (2010).

\bibitem{026-diaz-11}
J.S.\ Diaz, and V.A.\ Kosteleck\'y,
Phys.\ Lett.\ B {\bf 700}, 25 (2011).

\bibitem{026-diaz-12}
J.S.\ Diaz, and V.A.\ Kosteleck\'y,
Phys.\ Rev.\ D {\bf 85}, 016013 (2012).

\bibitem{026-barger-11}
V.\ Barger, D.\ Marfatia, and K.\ Whisnant,
Phys.\ Lett.\ B {\bf 653}, 267 (2007).

\bibitem{026-katori-11}
T.\ Katori, MiniBooNE Collaboration, in
V.A.\ Kosteleck\'y, ed.,
{\it CPT and Lorentz Symmetry V}
(World Scientific, Singapore, 2011).

\bibitem{026-bailey-06}
Q.G. Bailey and V.A.\ Kosteleck\'y,
Phys.\ Rev.\ D {\bf 74}, 045001 (2006).

\bibitem{026-battat-07}
J.B.R.\ Battat, J.F.\ Chandler, and C.W.\ Stubbs,
Phys.\ Rev.\ Lett.\ {\bf 99}, 241103 (2007).

\bibitem{026-chung-09}
K.-Y.\ Chung {\it et al.},
Phys.\ Rev.\ D {\bf 80}, 016002 (2009).

\bibitem{026-muller-08}
H.\ M\"uller {\it et al.},
Phys.\ Rev.\ Lett.\ {\bf 100}, 031101 (2008).

\bibitem{026-kostelecky-09b}
V.A.\ Kosteleck\'y and J.\ Tasson,
Phys.\ Rev.\ Lett.\ {\bf 102}, 010402 (2009).

\bibitem{026-kostelecky-11c}
V.A.\ Kosteleck\'y and J.\ Tasson,
Phys.\ Rev.\ D {\bf 83}, 016013 (2011).

\bibitem{026-honesee-11}
M.A.\ Hohensee {\it et al.},
Phys.\ Rev.\ Lett.\ {\bf 106}, 151102 (2011).

\bibitem{026-iorio-12}
L.\ Iorio,
Class.\ Quant.\ Gravit.\ {\bf 29}, 175007 (2012).

\bibitem{026-kostelecky-08b}
V.A.\ Kosteleck\'y, N.\ Russell, and J.\ Tasson,
Phys.\ Rev.\ Lett.\ {\bf 100}, 111102 (2008).

\bibitem{026-kostelecky-09c}
V.A.\ Kosteleck\'y and M.\ Mewes,
Phys.\ Rev.\ D {\bf 80}, 015020 (2009).

\bibitem{026-kostelecky-12a}
V.A.\ Kosteleck\'y and M.\ Mewes,
Phys.\ Rev.\ D {\bf 85}, 096005 (2012).

	

\end{thebibliography}
\end{document}